\documentstyle[12pt,aaspp4,flushrt]{article}

\def\spot{{\Large{$\bullet$~}}}
\slugcomment{Submitted to the Astrophysical Journal}
\lefthead{MIGHELL}
\righthead{PARAMETER ESTIMATION II.}
\begin{document}
\def\et{\hbox{et~al.\ }}
\def\paperi{\hbox{{\sc{PaperI}}}}
\def\apjlongeqi{\right. \nonumber \\ & & \left.}
\def\apjlongeqii{\right. \right. \nonumber \\ & & \left. \left.}
\def\lea{\mathrel{<\kern-1.0em\lower0.9ex\hbox{$\sim$}}}
\def\gea{\mathrel{>\kern-1.0em\lower0.9ex\hbox{$\sim$}}}
\def\ifundefined#1{\expandafter\ifx\csname#1\endcsname\relax}
\def\firstuse#1{\marginpar[]{{\scriptsize\hspace*{2.0mm}$\bullet${\sc{#1}}}}}
\def\noteforeditor#1{}

\title{
PARAMETER ESTIMATION IN ASTRONOMY\\
WITH POISSON-DISTRIBUTED DATA.\\
II. THE MODIFIED CHI-SQUARE-GAMMA STATISTIC
}

\author{
\sc
Kenneth J. Mighell
}
\affil{
\small
Kitt Peak National Observatory,\\
National Optical Astronomy Observatory\altaffilmark{1},\\
P. O. Box 26732, Tucson, AZ~~85726\\
Electronic mail:
mighell@noao.edu
\\[1cm]
}

\altaffiltext{1}{NOAO is operated
by the Association of Universities for Research in Astronomy, Inc., under
cooperative agreement with the National Science Foundation.}

\begin{abstract}
I investigate the use of
Pearson's chi-square statistic,
the Maximum Likelihood Ratio statistic for Poisson distributions,
and the chi-square-gamma statistic
(Mighell 1999, ApJ, 518, 380)
for the determination of the goodness-of-fit
between theoretical models and low-count Poisson-distributed data.
I demonstrate that these statistics should not be used to determine
the goodness-of-fit with data values of 10 or less.

I modify the chi-square-gamma statistic
for the purpose of improving its
goodness-of-fit performance.
I demonstrate that the
modified chi-square-gamma statistic
performs (nearly) like an ideal $\chi^2$ statistic
for the determination of goodness-of-fit with low-count data.
On average,
for correct (true) models,
the mean value of modified chi-square-gamma statistic
is equal to the number of degrees of freedom $(\nu)$
and its variance is $2\nu$
--- like the $\chi^2$ distribution for $\nu$ degrees of freedom.
Probabilities for modified chi-square-gamma goodness-of-fit values
can be calculated with the incomplete gamma function.

I give a practical demonstration showing how the
modified chi-square-gamma statistic can be used
in experimental astrophysics by analyzing
simulated X-ray observations of a weak point source
(S/N$\approx5.2\,$; 40 photons spread over 317 pixels)
on a noisy background
(0.06 photons per pixel).
Accurate estimates (95\% confidence intervals/limits)
of the location and intensity of the X-ray point source are determined.
\end{abstract}

\newpage
\section{INTRODUCTION}

The goodness-of-fit between an observation of $N$ data values, $x_i$,
with errors, $\sigma_i$, and a model, $m_i$,
can be determined by using the standard chi-square statistic:
\begin{equation}
\chi^2
\equiv
{\sum_{i=1}^{N}}
\left[
\frac{ x_i - m_i }{ \sigma_i }
\right]^2\,.
\label{eq:chisq_std}
\end{equation}
The standard chi-square statistic is the appropriate statistic to
determine the goodness-of-fit whenever the errors can be described by
a normal (a.k.a. Gaussian) distribution.

Let us consider a more complicated situation where
all the data values come from a pure counting
experiment where each measurement\footnote{
For example, X-ray photons, molecules, stars, galaxies, et cetera.
}, $n_i$, is a random integer deviate
drawn from a Poisson (\cite{po1837}, p.\ 205 et seq.)
distribution,
\begin{equation}
P(k;\mu)
\equiv
\frac{\mu^k}{k!}e^{-\mu}\,,
\label{eq:poisson}
\end{equation}
with a mean value of $\mu$.
The use of equation (\ref{eq:chisq_std}) in analyzing Poisson-distributed
data is technically never correct.
While the Poisson distribution approaches the normal distribution
as the Poisson mean approaches infinity, a Poisson distribution
never actually becomes a normal distribution even at very large
Poisson mean values.
The normal distribution is always symmetric;
the coefficient of skewness for the normal distribution is zero.
Poisson distributions are always asymmetric;
the coefficient of skewness for a Poisson distribution of mean $\mu$
is $\mu^{-1/2}$.
While the Poisson distribution is almost symmetric about the mean
for large mean values, its shape becomes
progressively more asymmetric as the mean approaches zero.
Thus the standard assumption that a Poisson distribution
is approximately normally distributed is a good approximation
only when the coefficient of skewness is negligible
(e.g.,\ $\mu^{-1/2}\!\ll\!1$).

How does one then determine the goodness-of-fit with Poisson-distributed data?
Historically, many $\chi^2$ statistics have been proposed
for the analysis of Poisson-distributed data.
This paper will investigate the following four:
\\[1.5em]
Pearson's $\chi^2$:
\begin{equation}
\chi_{\rm{P}}^2
\equiv
{\sum_{i=1}^{N}}
\frac{ (n_i - m_i)^2 }{ m_i }
\ ,
\label{eq:x2p}
\end{equation}
where the expectation value of the mean of the parent Poisson distribution
of the $i$th data value is assumed to be equal to the Poisson deviate
[$\langle\,\mu_i\,\rangle=n_i$]
and the square of the measurement error is
assumed to be equal to the mean of the model
Poisson distribution [$\sigma^2_i=m_i$];
\\[1.5em]
the modified Neyman's $\chi^2$:
\begin{equation}
\chi_{\rm{N}}^2
\equiv
{\sum_{i=1}^{N}}
\frac{ (n_i - m_i)^2 }{ {\rm{max}}(n_i,1) }
\ ,
\label{eq:x2n}
\end{equation}
where the expectation value of the mean of the parent Poisson distribution
of the $i$th data value is assumed to be equal to the Poisson deviate
[$\langle\,\mu_i\,\rangle=n_i$]
and the square of the measurement error is
assumed to be equal to Poisson deviate or one --- whichever is greater
[$\sigma^2_i={\rm{max}}(n_i,1)$];
\\[1.5em]
the Maximum Likelihood Ratio statistic for Poisson distributions:
\begin{equation}
\chi^2_\lambda
\equiv
2{\sum_{i=1}^{N}}
\left[
m_i - n_i + n_i\ln\left(\frac{n_i}{m_i}\right)
\right]
\label{eq:x2l}
\end{equation}
(see, e.g., Baker \& Cousins \cite{baco1984} and references therein);
\\[1.5em]
and the chi-square-gamma statistic
(Mighell \cite{mi1999}; hereafter \paperi):
\begin{equation}
\chi^2_\gamma
\equiv
{\sum_{i=1}^{N}}
\frac{ \left[ n_i + \min\left( n_i, 1\right) - m_i \right]^2 }{ n_i+1 }
\label{eq:x2g}
\ ,
\end{equation}
where the expectation value of the mean of the parent Poisson distribution
of the $i$th data value is assumed to be equal to
the Poisson deviate plus a small correction factor
of zero for zero deviates and one in all other cases
[$\langle\,\mu_i\,\rangle=n_i+\min(n_i,1)$]
and the square of the measurement error is
assumed to be equal to the Poisson deviate plus one
[$\sigma^2_i=n_i+1$].

In \paperi, I
demonstrated that the application of
the standard weighted mean formula,
$
\left[\sum_i {n_i \sigma^{-2}_i}\right]
/
\left[\sum_i {\sigma^{-2}_i}\right]
$,
to determine the weighted mean
of data, $n_i$, drawn from a Poisson distribution, will,
on average,
underestimate the true mean by $\sim$$1$ for all Poisson mean
values larger than $\sim$$3$
when the common assumption is made
that the error of the $i$th observation is
$\sigma_i = \max(\sqrt{n_i},1)$.
This small, but statistically significant offset,
explains the long-known observation that chi-square minimization techniques
which use the modified Neyman's $\chi^2$ statistic [eq.\ (\ref{eq:x2n})]
to compare Poisson-distributed data with model values, $m_i$, will
typically predict a total number of counts that
underestimates the actual total
by about $1$ count per bin
(see, e.g.,
Bevington \cite{be1969},
Wheaton \et \cite{weet1995}).

Based on my
finding that the weighted mean of data
drawn from a Poisson distribution can be
determined using the formula
$
\left[
\sum_i \left[n_i+\min\left(n_i,1\right)\right]\left(n_i+1\right)^{-1}
\right]
/
\left[
\sum_i \left(n_i+1\right)^{-1}
\right]
$,
I proposed that the chi-square-gamma statistic,
$\chi^2_\gamma$ [eq.\ (\ref{eq:x2g})],
should always be used to analyze Poisson-distributed data
in preference to the modified Neyman's $\chi^2$ statistic.
Following my own advice,
I will not discuss the modified Neyman's $\chi^2$ statistic
in the remainder of this article.

The chi-square distribution for $\nu$ degrees of freedom
approaches a Gaussian distribution with
a mean equal to $\nu$ (i.e., $\mu \equiv \nu$)
and a variance equal to $2\nu$ (i.e., $\sigma^2\equiv2\nu$)
as the number of degrees of freedom approaches infinity.
Ideally, a $\chi^2$ statistic for Poisson distributions for
$\nu$ (independent) degrees of freedom would exhibit the same behavior as the
number of degrees of freedom approaches infinity
{\em for all Poisson mean values}
(i.e.,\ $\mu\!>\!0$).

Do the
$\chi^2_{\rm{P}}$,
$\chi^2_\lambda$,
and the
$\chi^2_\gamma$
statistics perform as expected for large Poisson mean values?
These three $\chi^2$ statistics are applied to the same
data set in
Fig.\ \ref{fig:x2p_x2l_x2g_mu100}\firstuse{Fig\ref{fig:x2p_x2l_x2g_mu100}}
({\em{top to bottom}}, respectively).
For this example, an ideal $\chi^2$ statistic for Poisson-distributed data
would have a cumulative distribution similar to
that of the chi-square distribution for $10^4$ degrees of freedom
which well approximated as the cumulative distribution function
of a Gaussian distribution with a mean of $10^4$ and a variance of
$2\!\times\!10^4$.
The results of the top and bottom panels
are well matched to the expected cumulative distribution;
the differences between the expected
and measured mean and rms values are not statistically significant.
The $\chi^2_{\rm{P}}$ and the $\chi^2_\gamma$
statistics perform as expected with a Poisson mean value of 100.
The cumulative distribution of the middle panel, however,
clearly deviates from the expected cumulative distribution;
the difference between the expected and measured mean and rms values,
while small, is statistically significant.
The $\chi^2_\lambda$ does not perform like an ideal $\chi^2$ statistic
for Poisson distributions with a mean value of 100 --- a level
that is generally considered to be well above the low-count regime
($\mu\lea25$).

Let us continue the investigation of
the performance of the Maximum Likelihood Ratio
statistic for Poisson distributions with 1000 samples of $10^4$ Poisson
deviates with Poisson mean values of 100, 10, 1, 0.1, and 0.001.
Figure
\ref{fig:x2l_5mu}\firstuse{Fig\ref{fig:x2l_5mu}}
confirms that the $\chi^2_\lambda$ statistic does not perform
like an ideal $\chi^2$
statistic in the low-count regime.
The average contribution by the $i$th deviate
to an ideal $\chi^2$ statistic for the analysis of Poisson-distributed data
would be exactly one and the average contribution
to its variance would be exactly two.
Figure
\ref{fig:x2lr_var}\firstuse{Fig\ref{fig:x2lr_var}}
expands the previous analysis of \paperi\
of the $\chi^2_\lambda$ statistic
over a wide range of Poisson mean values
from 0.001 to 1000.
The dashed lines of Fig.\ \ref{fig:x2lr_var} show the
results for an ideal $\chi^2$ statistic; one can clearly see that
the average contribution to $\chi^2_\lambda$
is not equal to one
and the average contribution to its variance is not equal to two
for Poisson mean values $\lea$10.
The poor performance of the $\chi^2_\lambda$ statistic with low-count data
may come as a surprise to many readers since it has
historically been advocated as being one of the best
$\chi^2$ statistics for the analysis of Poisson-distributed data.

Chi-square statistics can serve (at least) two distinct purposes:
(1) their functional forms can be utilized as the core of parameter
estimation algorithms,
and
(2) their values can serve as a measure of the goodness-of-fit
between a model and a data set.\newline
\spot
While the functional form of the $\chi^2_\lambda$ statistic
can be successfully utilized for the purpose of
parameter estimation with Poisson-distributed data
in the low-count regime {\rm(see, e.g., \paperi)},
{\em{the Maximum Likelihood Ratio statistic for Poisson distributions
[eq.\ (\ref{eq:x2l})]
should not be used to determine the goodness-of-fit with low-count data
where the Poisson mean is $\lea$10.}}

In this work,
I investigate the use of
Pearson's $\chi^2$ statistic and the chi-square-gamma statistic
for the determination of the goodness-of-fit
between theoretical models and
data derived from counting experiments.
I develop a methodology in \S 2 which modifies
Pearsons's chi-square statistic for the purpose of improving its
goodness-of-fit performance.
This methodology is then be applied to modify the
chi-square-gamma statistic (\S 3).
The modified chi-square-gamma statistic is shown to
perform (nearly) like an ideal $\chi^2$ statistic
for the determination of goodness-of-fit with low-count data.
Simulated X-ray images are analyzed in \S 4 as a
practical demonstration of the possible use of the modified chi-square-gamma
statistic in experimental astrophysics.
The summary of the paper is presented in \S 5.

\section{THE MODIFIED PEARSON'S $\chi^2$ STATISTIC}

Let us continue the investigation of the performance of Pearson's $\chi^2$
statistic with 1000 samples of $10^4$ Poisson
deviates with Poisson mean values of 100, 10, 1, 0.1, and 0.001.
Figure
\ref{fig:x2p_5mu}\firstuse{Fig\ref{fig:x2p_5mu}}
shows that the $\chi^2_{\rm{P}}$ statistic
does not perform like an ideal $\chi^2$ statistic
for Poisson mean values $\lea\!10$.
The average contribution by the $i$th deviate
to an ideal $\chi^2$ statistic for the analysis of Poisson-distributed data
would be exactly one and the average contribution
to its variance would be exactly two.
Figure
\ref{fig:x2pr_var}\firstuse{Fig\ref{fig:x2pr_var}}
expands the previous analysis of \paperi\
of Pearson's $\chi^2$ statistic
over a wide range of Poisson mean values
from 0.001 to 1000.
The dashed lines of Fig.\ \ref{fig:x2pr_var} show the
results for an ideal $\chi^2$ statistic; one can see that
while the average contribution to $\chi^2_{\rm{P}}$
is one,
the average contribution to its variance is not equal to two
for Poisson mean values $\lea\!10$.\newline
\spot{\em{Pearson's $\chi^2$ statistic [eq.\ (\ref{eq:x2p})]
should not be used to determine the goodness-of-fit with low-count data
where the mean of the parent Poisson distribution is $\lea\!10$.}}

The variance of Pearson's $\chi^2$ statistic is, by definition,
\begin{equation}
\sigma^2_{\chi^2_{\rm{P}}}
\equiv
{\sum_{i=1}^{N}}
\left[
\frac{ (n_i - m_i)^2 }{ m_i }
-
\left(
\frac{1}{\nu}
{\sum_{j=1}^{N}}
\frac{ (n_j - m_j)^2 }{ m_j }
\right)
\right]^2\,,
\end{equation}
where $\nu = N - M$ is the number of independent degrees of freedom,
N is the number of data values, and M is the number of free parameters.
The variance of the {\em{reduced}} chi-square of a $\chi^2$ statistic
for a large number of observations should ideally be two.
In the limit of a large number of observations
of a {\em{single Poisson distribution with a mean value of $\mu$}},
the variance of the reduced chi-square of the Pearson's $\chi^2$ statistic is
\begin{eqnarray}
\sigma^2_{\chi^2_{\rm{P}}/\infty}
&\equiv
&\lim_{N \rightarrow \infty}
\left[
\frac{\sigma^2_{\chi^2_{\rm{P}}}}{\nu}
\right]
\nonumber
\\
&\equiv
&\lim_{N \rightarrow \infty}
\left[
\frac{1}{\nu}~{\sum_{i=1}^{N}}
\left\{
\frac{ (n_i - m_i)^2 }{ m_i }
-
\left[
\frac{1}{\nu}~{\sum_{j=1}^{N}}
\frac{ (n_j - m_j)^2 }{ m_j }
\right]
\right\}^2
\right]
\nonumber
\\
&\equiv
&\lim_{N \rightarrow \infty}
\left[
\frac{1}{\nu}~{\sum_{i=1}^{N}}
\left\{
\frac{ (n_i - m_i)^2 }{ m_i }
-
\left[
\frac{\chi^2_{\rm{P}}}{\nu}
\right]
\right\}^2
\right]
\nonumber
\\
&\approx
&\lim_{N \rightarrow \infty}
\left[
\frac{1}{\nu}~{\sum_{i=1}^{N}}
\left\{
\frac{ (n_i - m_i)^2 }{ m_i }
-
\lim_{N \rightarrow \infty}
\left[
\frac{\chi^2_{\rm{P}}}{\nu}
\right]
\right\}^2
\right]
\nonumber
\\
&=
&
\lim_{N \rightarrow \infty}
\left[
\frac{1}{N-M}~{\sum_{i=1}^{N}}
\left\{
\frac{ (n_i - m_i)^2 }{ m_i }
-
1
\right\}^2
\right]
\hspace*{22truemm}
\mbox{[see eq.\ (25) of \paperi]}
\nonumber
\\
&=
&\lim_{N \rightarrow \infty}
\left[
\frac{1}{N-1}~{\sum_{i=1}^{N}}
\left\{
\frac{ (n_i - \mu_{\rm{P}})^2 }{ \mu_{\rm{P}} }
-
1
\right\}^2
\right]
\hspace*{25truemm}
\mbox{[see eq.\ (5) of \paperi]}
\nonumber
\\
&\approx
&\lim_{N \rightarrow \infty}
\left[
\frac{1}{N-1}~{\sum_{i=1}^{N}}
\left\{
\frac{ \left(n_i - {\displaystyle\lim_{N \rightarrow \infty}}
[\mu_{\rm{P}}]\right)^2
}{
{\displaystyle\lim_{N \rightarrow \infty}} [\mu_{\rm{P}}]
}
-
1
\right\}^2
\right]
\nonumber
\\
&=
&\lim_{N \rightarrow \infty}
\left[
\frac{1}{N-1}~{\sum_{i=1}^{N}}
\left\{
\frac{ (n_i - \mu)^2
}{
\mu
}
-
1
\right\}^2
\right]
\hspace*{27truemm}
\mbox{[see eq.\ (7) of \paperi]}
\nonumber
\\
&\approx
&\lim_{N \rightarrow \infty}
\left[
\frac{1}{N-1}~{\sum_{k=0}^{\infty}}
\left\{
N P(k;\mu)
\right\}
\left\{
\frac{ (k - \mu)^2
}{
\mu
}
-
1
\right\}^2
\right]
\nonumber
\\
&=
&
\frac{1}{\mu^2}~{\sum_{k=0}^{\infty}}
P(k;\mu)
\left[
(k-\mu)^2
-
\mu
\right]^2
\nonumber
\\
&=
&2 + \frac{1}{\mu}
\ .
\label{eq:x2pr_var}
\end{eqnarray}

If we assume that Pearson's $\chi^2$ applied to a large number of observations
of a single Poisson distribution with a mean value of $\mu$
always produces a normal distribution with a
mean equal to the number of degrees-of-freedom ($\nu$)
[see eq.\ (25) of \paperi]
and a variance of $\nu(2 + \mu^{-1})$
[see eq.\ (\ref{eq:x2pr_var})],
we can then attempt to create an ideal $\chi^2$ statistic for the analysis
of Poisson-distributed data by modifying Pearson's $\chi^2$ as follows:
\begin{equation}
\chi_{\rm{PM}}^2
\equiv
{\sum_{i=1}^{N}}
\left[
\left\{
  {\chi^2_{{\rm{P}}{i}}}
  -
  {\langle\chi^2_{{\rm{P}}i}\rangle}
\right\}
\left[
{\frac{
  2
}{
  \sigma^2_{\langle\chi^2_{{\rm{P}}i}\rangle}
}}
\right]^{1/2}
\!\!
+
1
\,
\right]
\label{eq:x2pm}
\end{equation}
where
\begin{equation}
  {\chi^2_{{\rm{P}}i}}
  \equiv
  \frac{ (n_i - m_i)^2 }{ m_i }
\end{equation}
is
the contribution of the $i$th data value to Pearson's $\chi^2$,
\begin{equation}
  {\langle\chi^2_{{\rm{P}}i}\rangle}
  \equiv
  1
  \label{eq:expectation_value_x2p}
\end{equation}
is the expectation value of
${\chi^2_{{\rm{P}}i}}$
[see eq.\ (25) of \paperi],
\begin{equation}
  \sigma^2_{\langle\chi^2_{{\rm{P}}i}\rangle}
  \equiv
  {2 + m_i^{-1}}
\end{equation}
is the variance of
${\langle\chi^2_{{\rm{P}}i}\rangle}$
[see eq.\ (\ref{eq:x2pr_var})].
Translating the mathematical notation to English, we have
(1) shifted the mean of the standard $\chi^2_{\rm{P}}$
distribution
from
$\nu$ times equation (\ref{eq:expectation_value_x2p})
to
zero,
(2) forced the variance of the shifted distribution to be exactly $2\nu$,
and then
(3) shifted the mean of the variance-corrected distribution from zero
back to $\nu$.
Thus, {\em{by definition}},
the modified Pearson's chi-square statistic ($\chi^2_{\rm{PM}}$) will
have a mean value of $\nu$ and a variance of $2\nu$
--- in the limit of a large number of observations.

Let us now investigate the performance of the modified Pearson's $\chi^2$
statistic with 1000 samples of $10^4$ Poisson
deviates with Poisson mean values of 100, 10, 1, 0.1, and 0.001.
Figure
\ref{fig:x2pmt_5mu}\firstuse{Fig\ref{fig:x2pmt_5mu}}
shows that $\chi^2_{\rm{PM}}$ results are significantly better
than $\chi^2_{\rm{P}}$ results
[Fig.\ \protect\ref{fig:x2p_5mu}]
--- especially for Poisson mean values less than 10.
Figure
\ref{fig:x2pmtr_var}\firstuse{Fig\ref{fig:x2pmtr_var}}
investigates the performance of the
modified Pearson's $\chi^2$ statistic
over a wide range of Poisson mean values
from 0.001 to 1000.
The dashed lines of Fig.\ \ref{fig:x2pmtr_var} show the
results for an ideal $\chi^2$ statistic; one can see that
while the average contribution to $\chi^2_{\rm{PM}}$
is 1, as expected, and
the average contribution to its variance is equal to 2, as expected,
the performance is not uniform for all Poisson mean values
--- the spread seen in the variance plot ({\em{bottom panel}})
increases as the Poisson mean approaches zero.

Figures
\ref{fig:x2pmt_5mu}
and
\ref{fig:x2pmtr_var}
indicate the the
modified Pearson's $\chi^2$ statistic
works well in the perfect case where one has {\em{a priori}} knowledge
of the true Poisson mean.
In a real experiment, the true mean of the parent Poisson distribution
is rarely (if ever) known and model parameters must be estimated from the
observations.
How well does the
modified Pearson's $\chi^2$ statistic
work with reasonable parameter estimates?
Comparing
Fig.\ \ref{fig:x2pms_5mu}\firstuse{Fig\ref{fig:x2pms_5mu}}
with
Fig.\ \ref{fig:x2pmt_5mu}
and
Fig.\ \ref{fig:x2pmsr_var}\firstuse{Fig\ref{fig:x2pmsr_var}}
with
Fig.\ \ref{fig:x2pmtr_var},
we see that the {\em{variances are significantly smaller}}
when a {\em realistic} model (i.e., the sample mean) is used
instead of a {\em perfect} model (i.e., the true mean).
A statistic that fails with reasonable
parameter estimates is not a very useful statistic
for the analysis of real observations with low-count data.
\newline
\spot
{\em{The modified Pearson's $\chi^2$ statistic [eq.\ (\ref{eq:x2pm})]
should not be used to determine the goodness-of-fit with low-count data
where the mean of the parent Poisson distribution is $\lea\!10$.}}

\section{THE MODIFIED CHI-SQUARE-GAMMA STATISTIC}

Let us continue the investigation of
the performance of the chi-square-gamma
statistic with 1000 samples of $10^4$ Poisson
deviates with Poisson mean values of 100, 10, 1, 0.1, and 0.001.
Figure
\ref{fig:x2gt_5mu}\firstuse{Fig\ref{fig:x2gt_5mu}}
confirms that the $\chi^2_\gamma$ statistic does not perform
like an ideal $\chi^2$
statistic in the low-count regime.
Figure
\ref{fig:x2gt_var}\firstuse{Fig\ref{fig:x2gt_var}}
expands the previous analysis of \paperi\
of the $\chi^2_\gamma$ statistic
over a wide range of Poisson mean values
from 0.001 to 1000.
The $\chi^2_\gamma$ statistic clearly
does not perform like an ideal $\chi^2$ statistic
for Poisson mean values $\lea\!10$.\newline
\spot{\em{The chi-square-gamma statistic [eq.\ (\ref{eq:x2g})]
should not be used to determine the goodness-of-fit with low-count data
where the mean of the parent Poisson distribution is $\lea\!10$.}}

The variance of the chi-square-gamma statistic is, by definition,
\begin{equation}
\sigma^2_{\chi^2_\gamma}
\equiv
{\sum_{i=1}^{N}}
\left[
\frac{ [n_i + {\rm{min}}(n_i,1) - m_i]^2 }{ n_i + 1 }
-
\left(
\frac{1}{\nu}
{\sum_{j=1}^{N}}
\frac{ [n_j + {\rm{min}}(n_j,1) - m_j]^2 }{ n_j + 1 }
\right)
\right]^2\,,
\end{equation}
where $\nu = N - M$ is the number of independent degrees of freedom,
N is the number of data values, and M is the number of free parameters.
The variance of the {\em{reduced}} chi-square of a $\chi^2$ statistic
for a large number of observations should ideally be two.
In the limit of a large number of observations
of a {\em{single Poisson distribution with a mean value of $\mu$}},
the variance of the reduced chi-square of the chi-square-gamma statistic is
\begin{eqnarray}
\sigma^2_{\chi^2_{\gamma}/\infty}
&\equiv
&\lim_{N \rightarrow \infty}
\left[
\frac{\sigma^2_{\chi^2_{\gamma}}}{\nu}
\right]
\nonumber
\\
&\equiv
&\lim_{N \rightarrow \infty}
\left[
\frac{1}{\nu}~{\sum_{i=1}^{N}}
\left\{
\frac{ \left[ n_i + {\rm{min}}(n_i,1) - m_i \right]^2 }{ n_i + 1 }
-
\left[
\frac{1}{\nu}~{\sum_{j=1}^{N}}
\frac{ \left[ n_j + {\rm{min}}(n_j,1) - m_j \right]^2 }{ n_j + 1 }
\right]
\right\}^2
\right]
\nonumber
\\
&\equiv
&\lim_{N \rightarrow \infty}
\left[
\frac{1}{\nu}~{\sum_{i=1}^{N}}
\left\{
\frac{ \left[ n_i + {\rm{min}}(n_i,1) - m_i \right]^2 }{ n_i + 1 }
-
\left[
\frac{\chi^2_{\gamma}}{\nu}
\right]
\right\}^2
\right]
\nonumber
\\
&\approx
&\lim_{N \rightarrow \infty}
\left[
\frac{1}{\nu}~{\sum_{i=1}^{N}}
\left\{
\frac{ \left[ n_i + {\rm{min}}(n_i,1) - m_i \right]^2 }{ n_i + 1 }
-
\lim_{N \rightarrow \infty}
\left[
\frac{\chi^2_{\gamma}}{\nu}
\right]
\right\}^2
\right]
\nonumber
\\
&=
&
\lim_{N \rightarrow \infty}
\left[
\frac{1}{N-M}~{\sum_{i=1}^{N}}
\left\{
\frac{ \left[ n_i + {\rm{min}}(n_i,1) - m_i \right]^2 }{ n_i + 1 }
-
\left[ 1 + e^{-\mu}\left(\mu-1\right) \right]
\right\}^2
\right]
\hspace*{1truemm}
\mbox{[see eq.\ (29) of \paperi]}
\nonumber
\\
&=
&\lim_{N \rightarrow \infty}
\left[
\frac{1}{N-1}~{\sum_{i=1}^{N}}
\left\{
\frac{ \left[ n_i + {\rm{min}}(n_i,1) - \mu_{\gamma} \right]^2 }{ n_i + 1 }
-
\left[ 1 + e^{-\mu}\left(\mu-1\right) \right]
\right\}^2
\right]
\hspace*{4truemm}
\mbox{[see eq.\ (18) of \paperi]}
\nonumber
\\
&\approx
&\lim_{N \rightarrow \infty}
\left[
\frac{1}{N-1}~{\sum_{i=1}^{N}}
\left\{
\frac{ \left[ n_i + {\rm{min}}(n_i,1)
-
{\displaystyle\lim_{N \rightarrow \infty}} [\mu_{\gamma}]
\right]^2
}{
n_i + 1
}
-
\left[ 1 + e^{-\mu}\left(\mu-1\right) \right]
\right\}^2
\right]
\nonumber
\\
&=
&\lim_{N \rightarrow \infty}
\left[
\frac{1}{N-1}~{\sum_{i=1}^{N}}
\left\{
\frac{ \left[ n_i + {\rm{min}}(n_i,1) - \mu \right]^2
}{
n_i + 1
}
-
\left[ 1 + e^{-\mu}\left(\mu-1\right) \right]
\right\}^2
\right]
\hspace*{6truemm}
\mbox{[see eq.\ (19) of \paperi]}
\nonumber
\\
&\approx
&\lim_{N \rightarrow \infty}
\left[
\frac{1}{N-1}~{\sum_{k=0}^{\infty}}
\left\{
N P(k;\mu)
\right\}
\left\{
\frac{ \left[ k + {\rm{min}}(k,1) - \mu \right]^2
}{
k + 1
}
-
\left[ 1 + e^{-\mu}\left(\mu-1\right) \right]
\right\}^2
\right]
\nonumber
\\
&=
&
{\sum_{k=0}^{\infty}}
P(k;\mu)
\left\{
\frac{ \left[ k + {\rm{min}}(k,1) - \mu \right]^2
}{
k + 1
}
-
\left[ 1 + e^{-\mu}\left(\mu-1\right) \right]
\right\}^2
\nonumber
\\
&=
&
\mu^3 e^{-\mu}\left[
  {\rm{Ei}}(\mu) - \gamma_{\rm{EM}} - {\rm{ln}}(\mu) + 4
\right]
-\mu^2
-\mu
+e^{-\mu}\!\left[ -2\mu^2 + 2\mu + 1 \right]
+e^{-2\mu}\!\left[ -\mu^2 + 2\mu - 1 \right]\!,
\label{eq:x2gr_var}
\end{eqnarray}
where
Ei$(x)$ is the exponential integral of $x$
$[
\mbox{Ei}(x)
=
-\int_{-x}^{\infty} \frac{e^{-t}}{t}\,dt
=
\int_{-\infty}^{x} \frac{e^{-t}}{t}\,dt
\mbox{~for~} x\!>\!0
]$
and
$\gamma_{\rm{EM}}
\equiv
\lim_{n \rightarrow \infty}
\left[
\left\{
\sum_{i=1}^{n}
\frac{1}{n}
\right\}
- \ln(n)
\right]
\approx
0.5772156649
$
is the Euler-Mascheroni constant.
Equation (\ref{eq:x2gr_var})
approaches the expected value of 2 for
large Poisson mean values
[see the solid curve in the bottom panel of Fig.\ \ref{fig:x2gt_var}].

\def\x2gmx{{\chi^2_{{\gamma}i}}}
\def\rx2g{{\langle \chi^2_{\gamma{i}} \rangle}}
\def\vrx2g{
m_i^3 e^{-m_i}\left[ {\rm{Ei}}(m_i) - \gamma - {\rm{ln}}(m_i) + 4 \right]
-m_i^2
-m_i
+e^{-m_i}\left[ -2m_i^2 + 2m_i + 1 \right]
+e^{-2m_i}\left[ -m_i^2 + 2m_i - 1 \right]
}
If we assume that chi-square-gamma statistic is
applied to a large number of observations
of a single Poisson distribution with a mean value of $\mu$
always produces a normal distribution with a
mean equal to the number of degrees of freedom ($\nu$)
times see equation (29) of \paperi\ and a variance of
$\nu$ times equation (\ref{eq:x2gr_var}),
we can then attempt to create an ideal $\chi^2$ statistic for the analysis
of Poisson-distributed data by modifying
the chi-square-gamma statistic as follows:
\begin{equation}
\chi_{\rm{\gamma{M}}}^2
\equiv
{\sum_{i=1}^{N}}
\left[
\left\{
\x2gmx
-
\rx2g
\right\}
\left[
{\frac{2}{\sigma^2_\rx2g}}
\right]^{1/2}
\!\!
+
1
\,
\right]
\label{eq:x2gm}
\end{equation}
where
\begin{equation}
\x2gmx
\equiv
\frac{
\left[ n_i + \min\left( n_i, 1\right) - m_i \right]^2
}{
n_i+1
}
\label{eq:ith_contribution_x2g}
\end{equation}
is
the contribution of the $i$th data value to the chi-square gamma statistic,
\begin{equation}
  \rx2g
  \equiv
  1 + e^{-m_i}\left(m_i - 1\right)
  \label{eq:expectation_value_x2g}
\end{equation}
is the expectation value of
$\x2gmx$
[see equation (29) of \paperi],
and
\begin{eqnarray}
  \sigma^2_\rx2g
  &\equiv
  & m_i^3 e^{-m_i}\left[ {\rm{Ei}}(m_i) - \gamma_{\rm{EM}}
    - {\rm{ln}}(m_i) + 4 \right]
    -m_i^2
    -m_i + \nonumber
  \\
  &
  & e^{-m_i}\left[ -2m_i^2 + 2m_i + 1 \right]
    + e^{-2m_i}\left[ -m_i^2 + 2m_i - 1 \right]
\end{eqnarray}
is the variance of
$\rx2g$
[see equation (\ref{eq:x2gr_var})].
Translating the mathematical notation to English, we have
(1) shifted the mean of the standard $\chi^2_{\rm{P}}$ distribution
from
$\nu$ times equation (\ref{eq:expectation_value_x2g})
to
zero,
(2) forced the variance of the shifted distribution to be exactly $2\nu$,
and then
(3) shifted the mean of the variance-corrected distribution from zero
back to $\nu$.
Thus, {\em{by definition}},
the modified chi-square statistic statistic ($\chi^2_{\gamma\rm{M}}$) will
have a mean value of $\nu$ and a variance of $2\nu$
--- in the limit of a large number of observations.

Let us now investigate the performance of the modified chi-square-gamma
statistic with 1000 samples of $10^4$ Poisson
deviates with Poisson mean values of 100, 10, 1, 0.1, and 0.001.
Figure
\ref{fig:x2gmt_5mu}\firstuse{Fig\ref{fig:x2gmt_5mu}}
shows that $\chi^2_{\gamma\rm{M}}$ results are significantly better
than the $\chi^2_\gamma$ results
[Fig.\ \protect\ref{fig:x2gt_5mu}]
--- especially for Poisson mean values less than 10.
Figure
\ref{fig:x2gmt_var}\firstuse{Fig\ref{fig:x2gmt_var}}
investigates the performance of the
$\chi^2_{\gamma\rm{M}}$
statistic over a wide range of Poisson mean values
from 0.001 to 1000.
The dashed lines of Fig.\ \ref{fig:x2gmt_var} show the
results for an ideal $\chi^2$ statistic; one can see that
while the average contribution to $\chi^2_{\gamma\rm{M}}$
is 1, as expected, and
the average contribution to its variance is equal to 2, as expected,
the performance is not uniform for all Poisson mean values.
The bump seen in the bottom panel of Fig.\ \ref{fig:x2gmt_var}
near the Poisson mean value of 10 is an artifact caused by
the $\rm{min}(n_i,1)$ offset in the numerator of the
definition of the chi-square-gamma statistic
[eq.\ (\ref{eq:x2g})].

Figures
\ref{fig:x2gmt_5mu}
and
\ref{fig:x2gmt_var}
indicate the the
modified chi-square-gamma statistic
works well in the perfect case where one has {\em{a priori}} knowledge
of the true Poisson mean.
How well does the
modified chi-square-gamma statistic
work with reasonable parameter estimates?
Comparing
Fig.\ \ref{fig:x2gms_5mu}\firstuse{Fig\ref{fig:x2gms_5mu}}
with
Fig.\ \ref{fig:x2gmt_5mu}
and
Fig.\ \ref{fig:x2gms_var}\firstuse{Fig\ref{fig:x2gms_var}}
with
Fig.\ \ref{fig:x2gmt_var},
we see that the results for the modified $\chi^2_\gamma$ statistic with
a realistic model (i.e., the sample mean)
are nearly identical\footnote{
The measured mean values appearing on the right side of top 3 panels of
Fig.\ \ref{fig:x2gms_5mu} are about 1 lower than the comparable value
given in Fig.\ \ref{fig:x2gmt_5mu}.
This is the result of losing
one degree-of-freedom due to the determination of the sample mean from
the data (i.e., $\nu$ drops from 10000 to 9999).
The scrambling caused by the modification of the $\chi^2_\gamma$ statistic
appears to have caused this expected loss of one degree-of-freedom
to vanish in the very-low-count data regime ($\mu\!\lea\!0.1$).
} to those obtained with
a perfect model (i.e., the true mean).\newline
\spot
{\em{The modified chi-square-gamma [eq.\ (\ref{eq:x2gm})]
statistic performs (nearly) like
an ideal $\chi^2$ statistic for the determination of the goodness-of-fit
with low-count data.  On average, for a large number of observations,
the mean value of $\chi^2_{\gamma\rm{M}}$ statistic
is equal to the number of degrees of freedom $(\nu)$
and its variance is $2\nu$
--- like the $\chi^2$ distribution for $\nu$ degrees of freedom.}}

\section{SIMULATED X-RAY IMAGES}

I now demonstrate the new modified chi-square-gamma statistic by
using it to study simulated X-ray images.
Cash (\cite{ca1979}) applied his $C$ statistic to the problem
of determining the position of a weak source in a X-ray image.
Let us use Cash's Point Spread Function (PSF) but
with a resolution of 100 pixels per unit area:
\begin{equation}
\phi(x,y)
\equiv
\left\{
\begin{array}{ll}
  \displaystyle
  \frac{\pi}{3} \left( 1 - r \right)
  &\mbox{for $r\leq1$,}
\\
  0
  &\mbox{for $r>1$,}
\end{array}
\right.
\label{eq:psf}
\end{equation}
where $r^2 = (x/10)^2 + (y/10)^2$.
This PSF has the volume integral of
\begin{equation}
\Phi(x,y)
\equiv
\left\{
\begin{array}{ll}
\displaystyle
6
\left[
\frac{r^2}{2}
-
\frac{r^3}{3}
\right]
&\mbox{for $r\leq1$,}
\\
  1
&\mbox{for $r>1$.}
\end{array}
\right.
\end{equation}
Figure \ref{fig:observation}\firstuse{Fig\ref{fig:observation}}
shows a simulated observation of a point source with an intensity of
40 X-ray photons on a background flux of 0.06 X-ray photons per pixel.
This observation contains 2786 pixels with 0 photons,
204 pixels with 1 photons, and 10 pixels with 2 photons.
There are a total of 56 photons found in the 317 pixels
within a radius of 10 pixels of the center of the
X-ray point source which is located
at the $(x,y)$ position of $(33,26)$.
This is clearly a marginal detection of a weak X-ray point
source on a noisy background;
the peak signal-to-noise ratio ($\sim$5.2) occurs at a radius of
$\sim$8 pixels.

We will now use the modified chi-square-gamma statistic to answer
the following questions about this X-ray image:\\[-10truemm]
\begin{enumerate}
\item
Is there an X-ray point source in the image?
\item
If so, where is it located?
\item
What is its total intensity?
\end{enumerate}
\vskip -2truemm
The exact determination of the location and intensity
of the X-ray point source in Fig.\ \ref{fig:observation}
is precluded by the fact that this particular observation
contains only low-count data ---
we must be content with realistic estimates
for the location and intensity
based on a detailed statistical analysis of the data.

Our first objective is to determine if there is an X-ray point source
in the observation.
One way this can be done is to investigate
the region(s) containing non-point-source pixels (data values).
This approach requires knowledge of the background flux level ---
which we will henceforth assume is constant throughout the entire image.
We begin by making a rough first estimate of the background flux
level by dividing the total number of photon in the image by the total
number of pixels: $0.0747$ $(\approx 224\,/\,3000)$ photons per pixel.

The background flux level estimate may be significantly improved
with a bit more work.
There are 18 photons in the 317 pixels within 10 pixels of the position
$(12,15)$ of Fig.\ \ref{fig:observation}.
Is the detection of 18 photons consistent with the expected value
of 23.6799 $(=0.0747\times317)$ photons?
The upper and lower 99.9\% single-sided confidence limits for 18 photons
are 35.35 and 7.662, respectively
[see Tables 1 and 2 of Gehrels \cite{ge1986}].
I conclude that the pixel at $(12,15)$ is a background pixel
(shown as such with a gray box in
Fig.\ \ref{fig:first_sky}\firstuse{Fig\ref{fig:first_sky}})
because the {\em expected} number of background photons lies within the range
of the upper and lower 99.9\% single-sided
confidence limits of the {\em observed} number
of photons (i.e., $7.662 \leq 23.6799 \leq 35.35$).
There are 56 photons in the 317 pixels within 10 pixels of the position
$(33,26)$ of Fig.\ \ref{fig:observation}.
The upper and lower 99.9\% single-sided confidence limits for 56 photons
are 83.1784 and 35.6834, respectively
[see eqs.\ (10) and (14) of Gehrels \cite{ge1986}].
I conclude that the pixel at $(33,26)$ is {\em not} a background pixel
because the expected number of background photons is less than the
lower 99.9\% single-sided confidence limit of the observed number
of photons (i.e., $23.6799 < 35.6834$).
This conclusion was expected since $(33,26)$ is the center of the X-ray source.
The 2676 gray pixels in
Fig.\ \ref{fig:first_sky}
have a total of 162 photons.
We can now make a second estimate of the background flux:
$0.0605$ $(\approx 162\,/\,2676)$ photons per pixel.
Repeating this process once more yields the final
estimate of the background flux:
$0.0601$ $(\approx 153\,/\,2547)$ photons per pixel.
The measurement error for this estimate is approximately
0.0049 $(\approx \sqrt{153+1}\,/\,2547)$.
The final estimate of the X-ray background flux,
$0.0601\!\pm\!0.0049$,
is in excellent agreement with the true value of 0.06
[see Fig.\ \ref{fig:final_sky}\firstuse{Fig\ref{fig:final_sky}}].

I conclude that Fig.\ \ref{fig:observation} has at least one X-ray
point source because the entire data set is not consistent with a X-ray
background flux of 0.0601 photons per pixel {\em for every pixel}.
Assuming that there is only one X-ray source, we can make the first rough
estimate of its location by stating that it probably is located
at a non-gray pixel location in
Fig.\ \ref{fig:final_sky}.
There are 453 non-background pixels
in Fig.\ \ref{fig:final_sky}
with a total of 71 photons.
This fact allows us to
restrict the uncertainty of the location of the
X-ray point source to about 15\% $(\approx453\,/\,3000)$ of the total image.
Assuming a background flux of 0.06 photons per pixel,
we expect that
27 $(\approx453\times0.06)$
photons of the total 71 photons found in the non-background pixels
would be due to the background and not the point source.
We can now make the first rough estimate of the intensity of the X-ray
source: 44 $(=\!71\!-\!27)$ photons.

There are 48 photons in the 317 pixels within 10 pixels of the position
$(24,26)$ of Fig.\ \ref{fig:observation}.
Is this photon sum consistent with a model of
a 40 photon point source {\em centered at that location}
on a background of 0.06 photons per pixel?
The upper and lower 95\% single-sided confidence limits for 48 photons is
61.05 and 37.20 photons,
respectively.
The model predicts that we should find 58.9802\footnote{
These computations only included pixels within an aperture if the center
of the pixel was within the given aperture radius; partial pixels whose
center was just outside aperture boundary were rejected.
The minimum number we would expect the model to predict is
58.8496 $[40 + (\pi\times10^2\times0.0600)]$ photons.
The maximum number we would expect the model to predict is
59.0200 $[40 + (317\times0.0600)]$ photons.
The model prediction lies within these extremes:
$58.8496 < 58.9802 < 59.0200$.
}
photons within a radius of 10 pixels.
The 48 photons found within 10 pixels
of $(24,26)$ is consistent with the model
(shown as such with a dark-gray circle in Fig.\ \ref{fig:final_sky}),
because the {\em expected} number of photons lies within the range
of the upper and lower 95\% single-sided
confidence limits of the {\em observed} number
of photons (i.e., $37.20 \leq 58.9802 \leq 61.05$).
There are 217 circled pixels in Fig.\ \ref{fig:final_sky}.
This fact allows us to further
restrict the uncertainty of the location of the
X-ray point source to $\sim$7.2\% $(\approx217\,/\,3000)$ of the total image.

Since the peak signal-to-noise ratio occurs near a radius of 8 pixels,
we now investigate if we can improve our estimate of the location of
the X-ray point source by considering photon sums within a smaller aperture
with a radius of 8 instead of 10 pixels.
There are 34 photons in the 197 pixels within a radius of
8 pixels of the position
$(26,26)$ of Fig.\ \ref{fig:observation}.
Is this photon sum consistent with a model of
a 40 photon point source centered at that location
with a background of 0.06 photons per pixel?
The upper and lower 95\% single-sided confidence limits for 34 photons is
45.27 and 25.01 photons,
respectively.
The model predicts that we should find 47.2664
photons within a radius of 8 pixels.
The 34 photons found within 8 pixels
of $(26,26)$ is {\em not} consistent with the model
since the expected number of photons is greater than the
upper 95\% single-sided confidence limit
of the observed number of photons
(i.e., $47.2664 > 45.27$).
However, the 39 photons found within 8 pixels of $(27,26)$
{\em is} consistent with a 40 photon point source centered at $(27,26)$
on a background of 0.06 photons per pixel
(i.e., $29.33 \leq 47.2664 \leq 50.94$).
There are a total of 111 circled pixels in
Fig.\ \ref{fig:green_radius8}
\firstuse{Fig\ref{fig:green_radius8}}.
This fact allows us to further
restrict the uncertainty of the location of the
X-ray point source to $\sim$3.7\% $(=111\,/\,3000)$ of the total image.

One way to boost the data out of the low-count regime is to compare
the cumulative radial distribution of the model with the cumulative
radial distribution of the data.
The modified chi-square-gamma statistic was used to compare
the cumulative radial distribution of the model
(10 1-pixel-wide bins $\Rightarrow$ 10 degrees-of-freedom)
with
the cumulative radial distribution of the data
(similarly formatted).
At the position of $(27,26)$ the value of $\chi^2_{\gamma{\rm M}}$
for the cumulative radial distributions
was computed to be 13.7338 ($\nu\equiv10$)
for a model of a 40 photon point source
centered at $(27,26)$ on a background of 0.06 photons per pixel.
The 95th percentage point for the chi-square distribution with 10
degrees of freedom may be found in several standard references:
18.3
({\em CRC Handbook of Chemistry and Physics},
Lide \& Frederikse \cite{crc1994}, p.\ A-106),
18.31
(Bevington \cite{be1969}, p.\ 315)
and
18.3070
(Abramowitz \& Stegun \cite{abst1964}, p.\ 985).
I conclude that the image location $(27,26)$ is within the
95\% confidence interval because the value of the
modified chi-square-gamma statistic
is less than the 95th percentage point
for the chi-square distribution with 10 degrees of freedom
(i.e. $13.7338\!<\!18.3070$).
The probability that the observed chi-square value
for a correct model should be less than a value of $\chi^2$ for
$\nu$ degrees of freedom is
$P(\chi^2|\nu) = P(\frac{\nu}{2},\frac{\chi^2}{2})$
where the latter function is the incomplete gamma function
[ $P\equiv1-Q$;
see, e.g., the {\tt{GAMMP}} routine in {\em Numerical Recipes}
(Press \et \cite{pret1986})].
If we assume that $\chi^2_{\gamma{\rm M}}$ is distributed like the $\chi^2$
distribution, then we can assign a probability for the
modified chi-square-gamma value for 10 degrees of freedom:
$P(13.7338|10) = P(\frac{10}{2},\frac{13.7338}{2})=0.814517$.
There is thus a
$\sim$81.5\% chance that the observed modified
chi-square-gamma statistic will be less than 13.7338 for 10 degrees
of freedom.
The contour in Fig.\ \ref{fig:green_radius8}
shows the 95\% confidence interval of the X-ray point source
based on the $\chi^2_{\gamma{\rm M}}$
analysis of the cumulative radial distribution of the data.
The value of $\chi^2_{\gamma{\rm M}}$
for the cumulative radial distribution at $(26,26)$
was computed to be 30.4707 giving
a probability of $\sim$99.9\%; this location
in Fig.\ \ref{fig:green_radius8} lies
outside the 95\% confidence interval.

We can further use the modified chi-square-gamma statistic with the
cumulative radial distribution to determine the 95\% confidence
limits of the intensity of the X-ray source in the image
(see
Fig.\ \ref{fig:crd_probability}
\firstuse{Fig\ref{fig:crd_probability}}
).
The upper and lower single-sided 95\% confidence limits for
the intensity of an X-ray point source at $(33,26)$ in
Fig.\ \protect\ref{fig:observation} is 54.5 and 28.0, respectively.
The true intensity of the X-ray source is 40 photons.
Given a background flux uncertainty of
$\sigma_B=0.0049$ photons per pixel
(see above), we can approximate the theoretical rms measurement error for
a 40 photon point source spread over 317 pixels $(A=317$ px$^2)$ as
$\sigma \approx \sqrt{40 + 1} + A\sigma_B \approx 8.0$ photons.
The difference between the upper and lower 95\% single-sided confidence
limits is approximately 3.3 standard deviations of the normal probability
function.
This fact
can be used to approximate an rms measurement error for our
intensity estimate of $\sigma_I\approx8.0$
[$\approx(54.5-28.0)/(2\times1.65)$]
photons.
The $\chi^2_{\gamma{\rm M}}$ analysis using the cumulative
radial distribution has yielded an excellent intensity estimate.

The analysis presented in Figures
\ref{fig:green_radius8}
and
\ref{fig:crd_probability}
is predicated on the assumption that
the modified chi-square-gamma statistic is distributed like $\chi^2$.
But is this assumption valid?
Figure \ref{fig:10000_cumulative_fraction}
\firstuse{Fig\ref{fig:10000_cumulative_fraction}}
shows that the analysis of $10^4$ simulated X-ray observations like
Fig.\ \ref{fig:observation}
yields modified chi-square-gamma values that are distributed
like the chi-square distribution for $\nu\equiv317$ degrees of freedom:
a Gaussian distribution with
a mean of $\nu$
and a variance of $2\nu$.
The above analysis has assumed that
the probability that the observed modified chi-square-gamma value
for a correct model should be less than a value of $\chi^2$ for
$\nu$ degrees of freedom can be given as
$P(\chi^2_{\gamma{\rm M}}|\nu)$.
Assuming that the predicted probability from
$P(\chi^2_{\gamma{\rm M}}|\nu)$
is an accurate prediction of the true
probability, then the {\em predicted} probability of the 9500th simulated
observation of a total of 10000
(sorted by $\chi^2_{\gamma{\rm M}}$ value) should be
very close to 95\%;
Fig.\ \ref{fig:10000_sorted_probability}
\firstuse{Fig\ref{fig:10000_sorted_probability}}
indicates that this is indeed the case
(i.e., the probability for the $\chi^2_{\gamma{\rm M}}$ value
of the 9500th simulated observation is 94.8666\%).
Since $\chi^2_{\gamma{\rm M}}$ statistic is distributed (nearly) like the
$\chi^2$ distribution, the usage of the incomplete gamma function
to predict probabilities for modified chi-square-gamma values
appears to be justified in practical analysis problems.

\section{SUMMARY}

I investigated the use of
Pearson's chi-square statistic
[eq.\ (\ref{eq:x2p})],
the Maximum Likelihood Ratio statistic for Poisson distributions
[eq.\ (\protect\ref{eq:x2l})],
and the chi-square-gamma statistic
[eq.\ (\protect\ref{eq:x2g})]
for the determination of the goodness-of-fit
between theoretical models and low-count Poisson-distributed data.
I concluded that none of these statistics should be used to determine
the goodness-of-fit with data values of 10 or less.

I modified Pearson's chi-square statistic
for the purpose of improving its
goodness-of-fit performance.
I demonstrated that
modified Pearson's $\chi^2$ statistic
[eq.\ (\ref{eq:x2pm})]
works well in the perfect case where one has {\em{a priori}} knowledge
of the correct (true) model.
In a real experiment, however,
the true mean of the parent Poisson distribution
is rarely (if ever) known and model parameters must be estimated from the
observations.
I demonstrated that the
modified Pearson's $\chi^2$ statistic
has a variance that is significantly smaller than that of the
$\chi^2$ distribution when {\em realistic} models,
defined as having parameters estimated from the observational data,
are compared with Poisson-distributed data.
Any statistic that fails with models based on reasonable
parameter estimates is not a very practical statistic for the
analysis of astrophysical observations.
I concluded that the
modified Pearson's $\chi^2$ statistic
should not be used to determine the goodness-of-fit with low-count data
values of 10 or less.

I modified the chi-square-gamma statistic
for the purpose of improving its
goodness-of-fit performance.
I demonstrated that the
modified chi-square-gamma statistic
[eq.\ (\ref{eq:x2gm})]
performs (nearly) like an ideal $\chi^2$ statistic
for the determination of goodness-of-fit with low-count data.
On average,
for correct (true) models,
the mean value of the modified chi-square-gamma statistic
is equal to the number of degrees of freedom $(\nu)$
and its variance is $2\nu$
--- like the $\chi^2$ distribution for $\nu$ degrees of freedom.

An ideal $\chi^2$ statistic
for the determination of goodness-of-fit with low-count data
should {\em fail} in a predictable manner.
Hypothesis testing of low-count Poisson-distributed data
with the modified Pearson's $\chi^2$ statistic
will produce the peculiar and undesirable result that
correct models are more likely to be rejected than realistic models
[cf.\ Fig.\ \ref{fig:x2pmt_5mu} with Fig.\ \ref{fig:x2pms_5mu}].
The modified chi-square-gamma statistic is a {\em practical} statistic to use
for hypothesis testing of astrophysical data from counting
experiments because
it performs (nearly) like an ideal $\chi^2$ statistic
for realistic {\em and} correct models in
the low-count {\em and} the high-count data regimes;
accurate and believable
probabilities for $\chi^2_{\gamma{\rm M}}$ goodness-of-fit values
can be calculated with the incomplete gamma function
[Figs.\ \ref{fig:10000_cumulative_fraction}
and
\ref{fig:10000_sorted_probability}].
A lot of nothing can tell you something --- as long as there are {\em some}
observations with signal in them.

\acknowledgments

Vincent Eke sent me an e-mail asking if I had an expression for
the variance of the $\chi^2_\gamma$ statistic which described the
mysterious second hump of the solid curve of Fig.\ 3 of \paperi.
After a rapid exchange of email with me over the period
of a week, he was the first to derive an analytical formula for
$\sigma^2_{\chi^2_{\gamma}/\infty}$
[eq.\ (\ref{eq:x2gr_var})].
The knowledge that the variance of $\chi^2_\gamma$ could in fact
be expressed explicitly
as an analytical expression turned out to be the breakthrough
that I had needed in order to
complete the development of the modified $\chi^2_\gamma$ statistic.
It is a pleasure to acknowledge his contribution to this research.

I would like to thank Mike Merrill for the use of his copy
of {\sl{Mathematica}} which I used to check some of the arithmetic
of the critical last step in the derivation of Eq.\ (\ref{eq:x2gr_var}).

Special thanks are due to Mary Guerrieri, the NOAO librarian,
who has greatly facilitated this research
effort by finding and securing loans for
many a quaint and curious volume of forgotten lore.

I was supported by a grant from
the National Aeronautics and Space Administration (NASA),
Order No.\ S-67046-F, which was awarded by
the Long-Term Space Astrophysics Program (NRA 95-OSS-16).
This research has made use of
NASA's Astrophysics Data System Abstract Service
which is operated by the Jet Propulsion Laboratory at
the California Institute of Technology,
under contract with NASA.

\clearpage
\mbox{~}
\clearpage

\clearpage
\def\fig01cap{
\label{fig:x2p_x2l_x2g_mu100}
\noteforeditor{Print this figure ONE (1) COLUMN wide.\newline}
A simulated data set of 1000 samples (``observations'') of
$10^4$ Poisson deviates (``measurements'') per sample
was created assuming a mean value $\mu$$\equiv$$100$ for each Poisson
deviate.
Each sample in this data set was then analyzed using
Pearson's $\chi^2$ statistic
[{\em{top;}}\ definition:\ eq.\ (\protect\ref{eq:x2p})],
the Maximum Likelihood Ratio statistic for Poisson distributions
[{\em{middle;}}\ definition:\ eq.\ (\protect\ref{eq:x2l})],
and the chi-square-gamma statistic
[{\em{bottom;}}\ definition:\ eq.\ (\protect\ref{eq:x2g})].
The model of the $i$th deviate in each sample
was set to the true mean value
of parent Poisson distribution
(i.e., $m_i\!=\!\mu\!\equiv\!100$) and
the number of independent degrees-of-freedom was
therefore equal to the number of deviates per sample
(i.e.\ $\nu\!\equiv\!10^4$).
Compare the cumulative distribution for each statistic
with the cumulative distribution function of a Gaussian distribution
with a mean of $10^4$ and a variance of $2\!\times\!10^4$
[{\em{thick curve}} in each panel].
The number and error shown on the right side of each panel is the mean and
rms value of the 1000 samples shown in that panel; ideally these values
should be about $10000\!\pm\!141.4\,$.
}

\ifundefined{showfigs}{
  \newpage
  \centerline{{\Large\bf{Figure Captions}}}
  \smallskip
  \figcaption{\fig01cap}
}\else{
  \clearpage
  \newpage
  \begin{figure}
  \figurenum{1}
  \vspace*{-38truemm}
  \hspace*{+18truemm}
  \epsfxsize=6.5truein
  \epsfbox{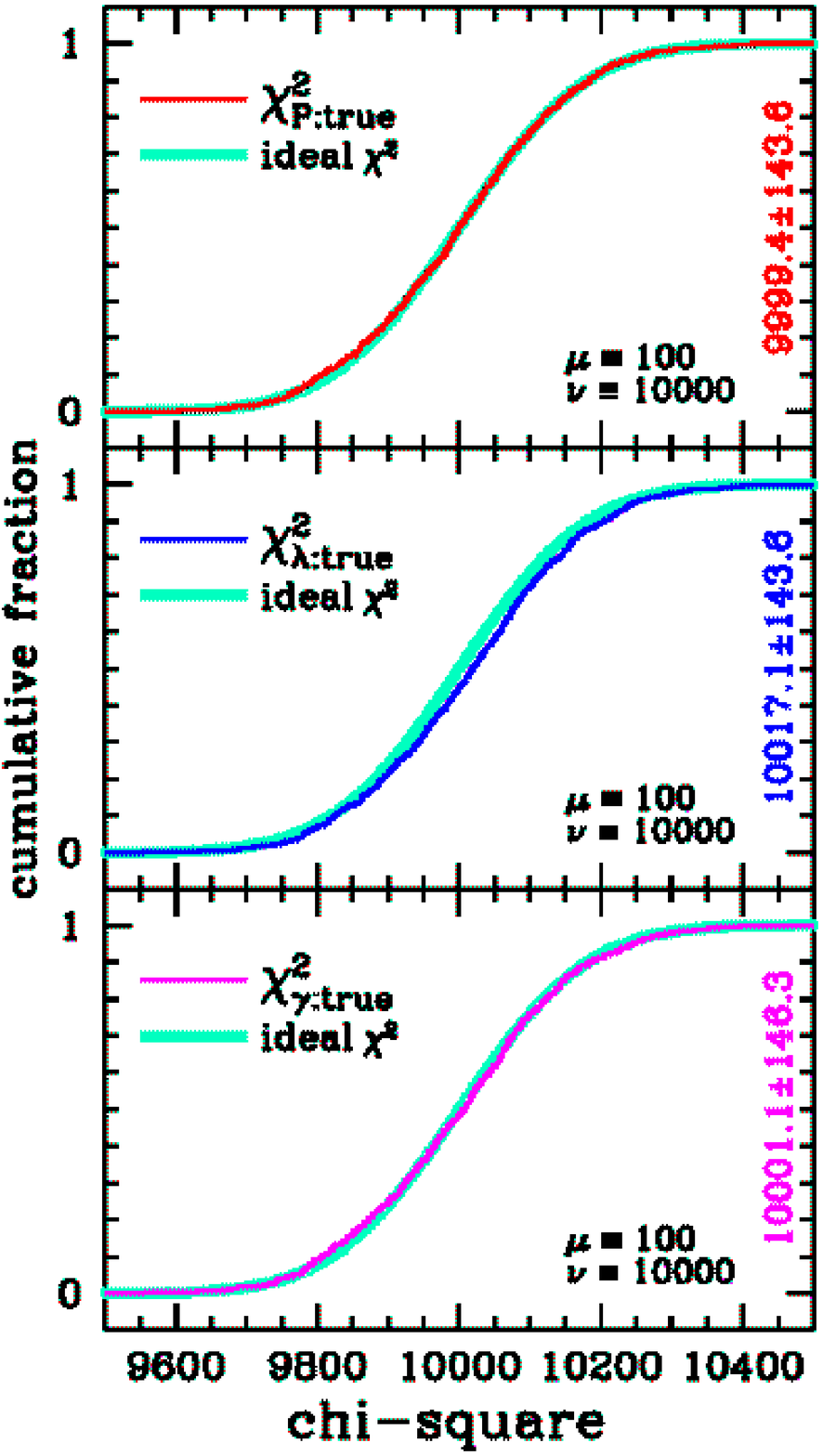}
  \vspace*{-40truemm}
  \caption[]{\baselineskip 1.15em \fig01cap}
  \end{figure}
}
\fi
\clearpage

\def\fig02cap{
\label{fig:x2l_5mu}
\noteforeditor{Print this figure ONE (1) COLUMN wide.\newline}
The cumulative distribution functions for
1000 samples of $10^4$ Poisson deviates
({\em{top to bottom}}: $\mu \equiv 100, 10, 1, 0.1$, and 0.01)
analyzed using
the Maximum Likelihood Ratio statistic for Poisson distributions
[definition:\ eq.\ (\protect\ref{eq:x2l})].
In all cases, $\nu\!\equiv\!10^4$ and $m_i$ was set to the true mean value
of the data set.
Other details as in Fig.\ \protect\ref{fig:x2p_x2l_x2g_mu100}.
}

\ifundefined{showfigs}{
  \figcaption{\fig02cap}
}\else{
  \clearpage
  \newpage
  \begin{figure}
  \figurenum{2}
  \vspace*{-40truemm}
  \hspace*{+18truemm}
  \epsfxsize=6.5truein
  \epsfbox{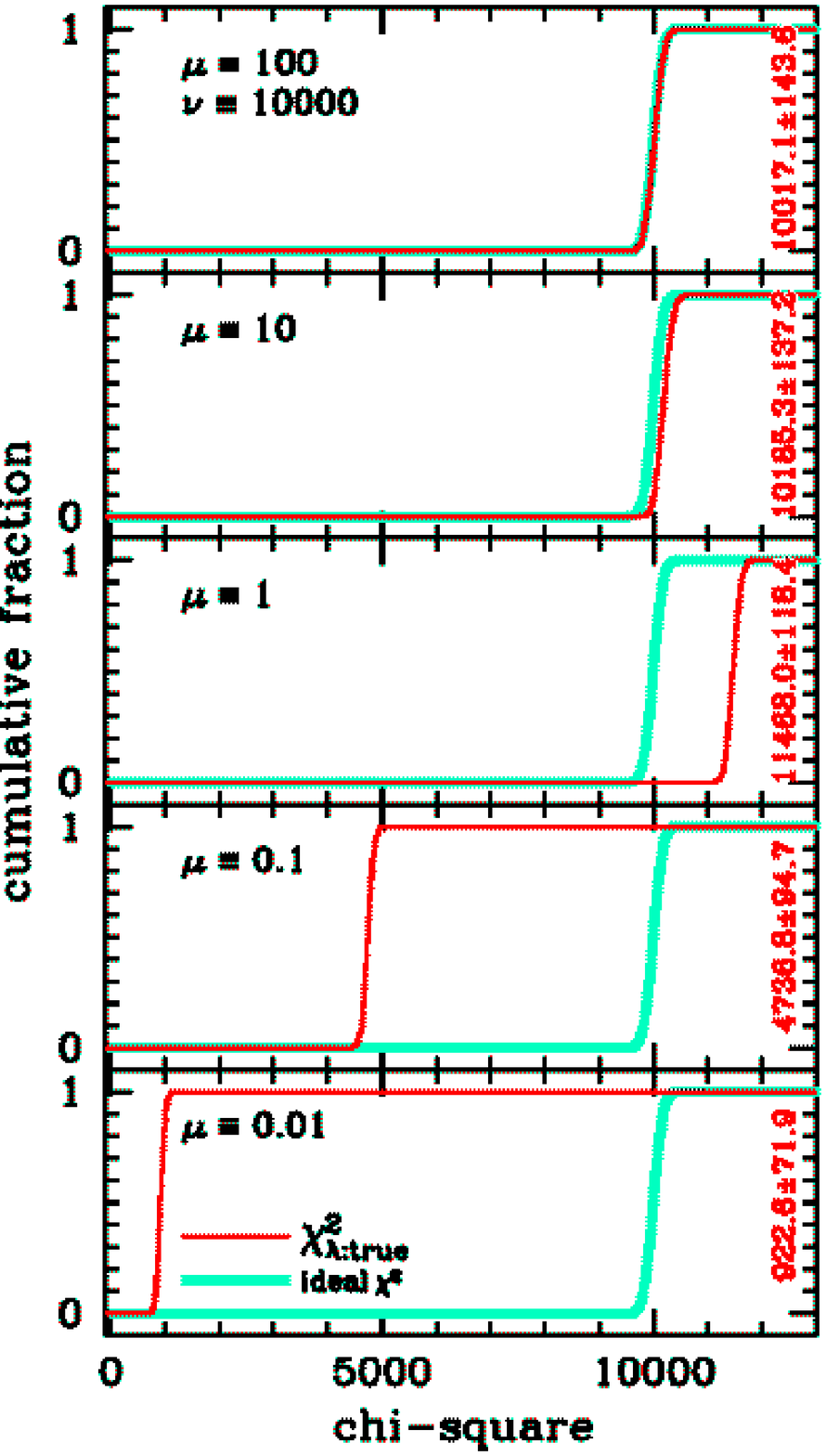}
  \vspace*{-40truemm}
  \caption[]{\baselineskip 1.15em \fig02cap}
  \end{figure}
}
\fi
\clearpage

\def\fig03cap{
\label{fig:x2lr_var}
\noteforeditor{Print this figure ONE (1) COLUMN wide.\newline}
{\em{Top panel:}}
Reduced chi-square
as a function of the true Poisson mean
($ 0.001 \leq \mu \leq 1000$ with 10 mean values per decade)
for the Maximum Likelihood Ratio statistic for Poisson distributions
with the model of the $i$th deviate set to the
mean value of the parent Poisson distribution.
The {\em{open squares}} show the results of the analysis of one
sample composed of $10^7$
Poisson deviates ($\nu\equiv10^7$) at each given Poisson mean value.
The {\em{filled squares}}
show the results of the analysis of 1000 subsamples of the $10^7$
Poisson deviates ($\nu\equiv10^4$) previously analyzed as one large sample.
The scatter of the filled squares with respect to the open squares
is real and is due to random fluctuations of the parent Poisson distributions.
The {\em{dashed line}} shows the ideal value of one.
{\em{Bottom panel:}}
The variance of the reduced chi-square values shown in the top panel.
The {\em{dashed line}} shows the ideal value of two.
}

\ifundefined{showfigs}{
  \figcaption{\fig03cap}
}\else{
  \clearpage
  \newpage
  \begin{figure}
  \figurenum{3}
  \vspace*{-40truemm}
  \hspace*{+18truemm}
  \epsfxsize=6.5truein
  \epsfbox{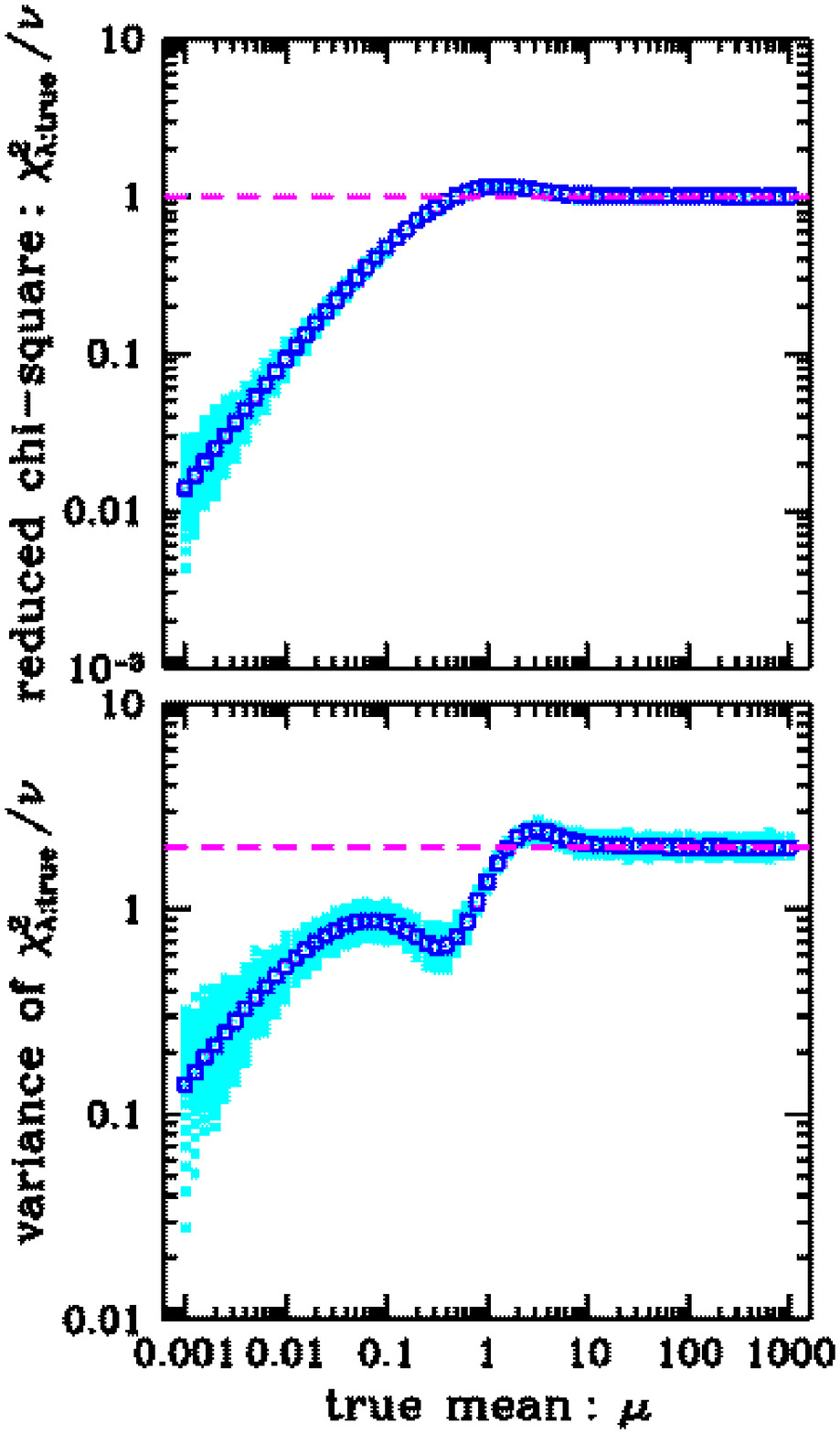}
  \vspace*{-40truemm}
  \caption[]{\baselineskip 1.15em \fig03cap}
  \end{figure}
}
\fi
\clearpage

\def\fig04cap{
\label{fig:x2p_5mu}
\noteforeditor{Print this figure ONE (1) COLUMN wide.\newline}
The cumulative distribution functions for
1000 samples of $10^4$ Poisson deviates
({\em{top to bottom}}: $\mu \equiv 100, 10, 1, 0.1$, and 0.01)
analyzed using Pearson's $\chi^2$ statistic
[definition:\ eq.\ (\protect\ref{eq:x2p})]
(same input data set as for Fig.\ \protect\ref{fig:x2l_5mu}).
In all cases, $\nu\!\equiv\!10^4$ and $m_i$ was set to the true mean value
of the data set.
Other details as in Fig.\ \protect\ref{fig:x2p_x2l_x2g_mu100}.
}

\ifundefined{showfigs}{
  \figcaption{\fig04cap}
}\else{
  \clearpage
  \newpage
  \begin{figure}
  \figurenum{4}
  \vspace*{-40truemm}
  \hspace*{+18truemm}
  \epsfxsize=6.5truein
  \epsfbox{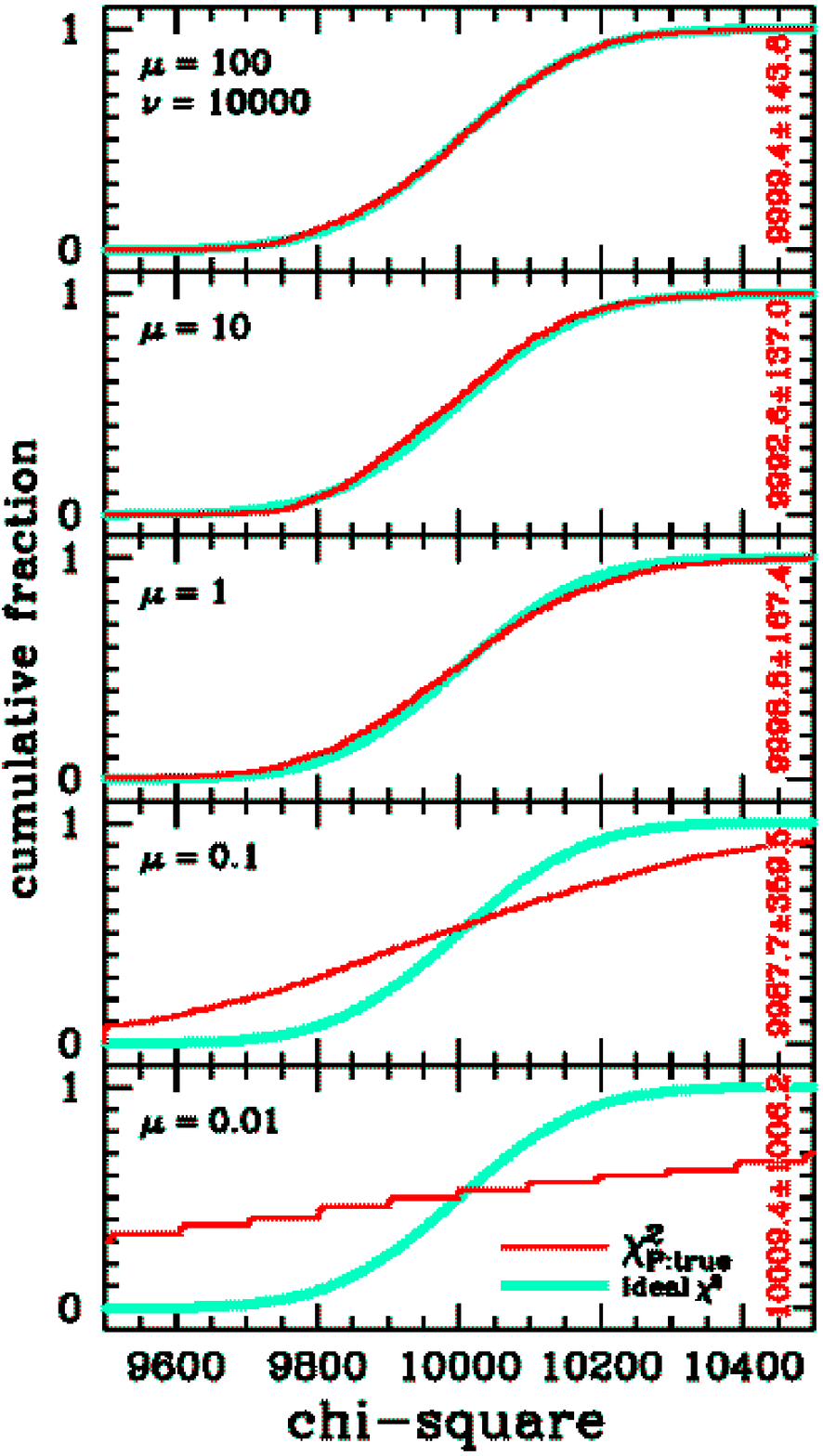}
  \vspace*{-40truemm}
  \caption[]{\baselineskip 1.15em \fig04cap}
  \end{figure}
}
\fi
\clearpage

\def\fig05cap{
\label{fig:x2pr_var}
\noteforeditor{Print this figure ONE (1) COLUMN wide.\newline}
Reduced chi-square as a function of the true Poisson mean for
Pearson's $\chi^2$ statistic
with the model of the $i$th deviate set to the
true mean value of the parent Poisson distribution
(same input data set as for Fig.\ \protect\ref{fig:x2lr_var}).
The {\em{solid line}} connecting the open squares in the
{\em{bottom panel}} is the formula
$2+\mu^{-1}$ [see eq.\ (\protect\ref{eq:x2pr_var})].
Other details as in Fig.\ \protect\ref{fig:x2lr_var}.
}

\ifundefined{showfigs}{
  \figcaption{\fig05cap}
}\else{
  \clearpage
  \newpage
  \begin{figure}
  \figurenum{5}
  \vspace*{-40truemm}
  \hspace*{+18truemm}
  \epsfxsize=6.5truein
  \epsfbox{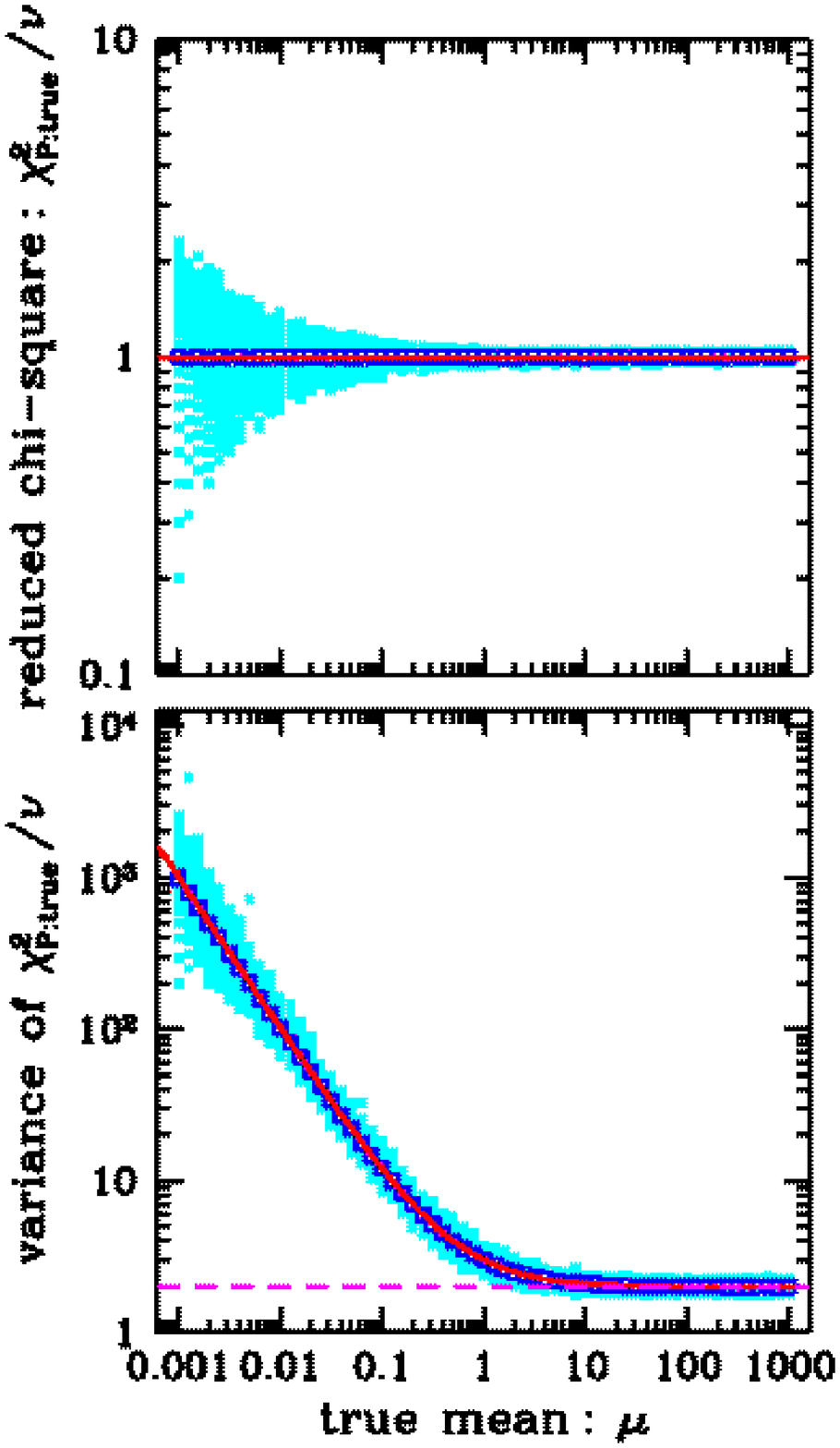}
  \vspace*{-40truemm}
  \caption[]{\baselineskip 1.15em \fig05cap}
  \end{figure}
}
\fi
\clearpage

\def\fig06cap{
\label{fig:x2pmt_5mu}
\noteforeditor{Print this figure ONE (1) COLUMN wide.\newline}
The cumulative distribution functions for
1000 samples of $10^4$ Poisson deviates
({\em{top to bottom}}: $\mu \equiv 100, 10, 1, 0.1$, and 0.01)
analyzed using the modified Pearson's $\chi^2$ statistic
[definition:\ eq.\ (\ref{eq:x2pm})]
(same input data set as for Fig.\ \protect\ref{fig:x2l_5mu}).
In all cases, $\nu\!\equiv\!10^4$ and $m_i$ was set to the true mean value
of the data set.
Other details as in Fig.\ \protect\ref{fig:x2p_x2l_x2g_mu100}.
}

\ifundefined{showfigs}{
  \figcaption{\fig06cap}
}\else{
  \clearpage
  \newpage
  \begin{figure}
  \figurenum{6}
  \vspace*{-40truemm}
  \hspace*{+18truemm}
  \epsfxsize=6.5truein
  \epsfbox{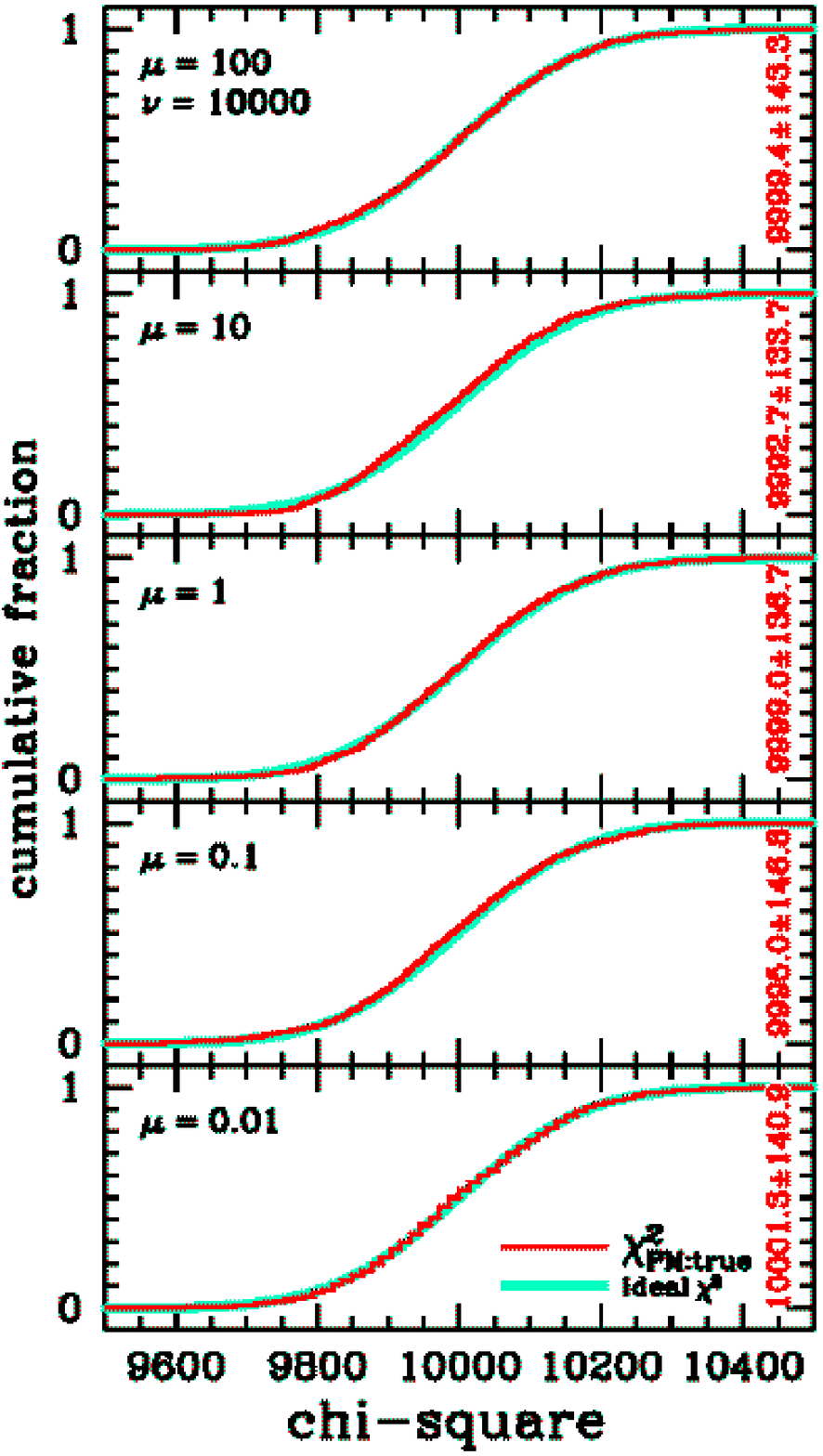}
  \vspace*{-40truemm}
  \caption[]{\baselineskip 1.15em \fig06cap}
  \end{figure}
}
\fi
\clearpage

\def\fig07cap{
\label{fig:x2pmtr_var}
\noteforeditor{Print this figure ONE (1) COLUMN wide.\newline}
Reduced chi-square as a function of the true Poisson mean for
the modified Pearson's $\chi^2$
statistic with the model of the $i$th deviate set to the
true mean value of the parent Poisson distribution
(same input data set as for Fig.\ \protect\ref{fig:x2lr_var}).
Other details as in Fig.\ \protect\ref{fig:x2lr_var}.
}

\ifundefined{showfigs}{
  \figcaption{\fig07cap}
}\else{
  \clearpage
  \newpage
  \begin{figure}
  \figurenum{7}
  \vspace*{-40truemm}
  \hspace*{+18truemm}
  \epsfxsize=6.5truein
  \epsfbox{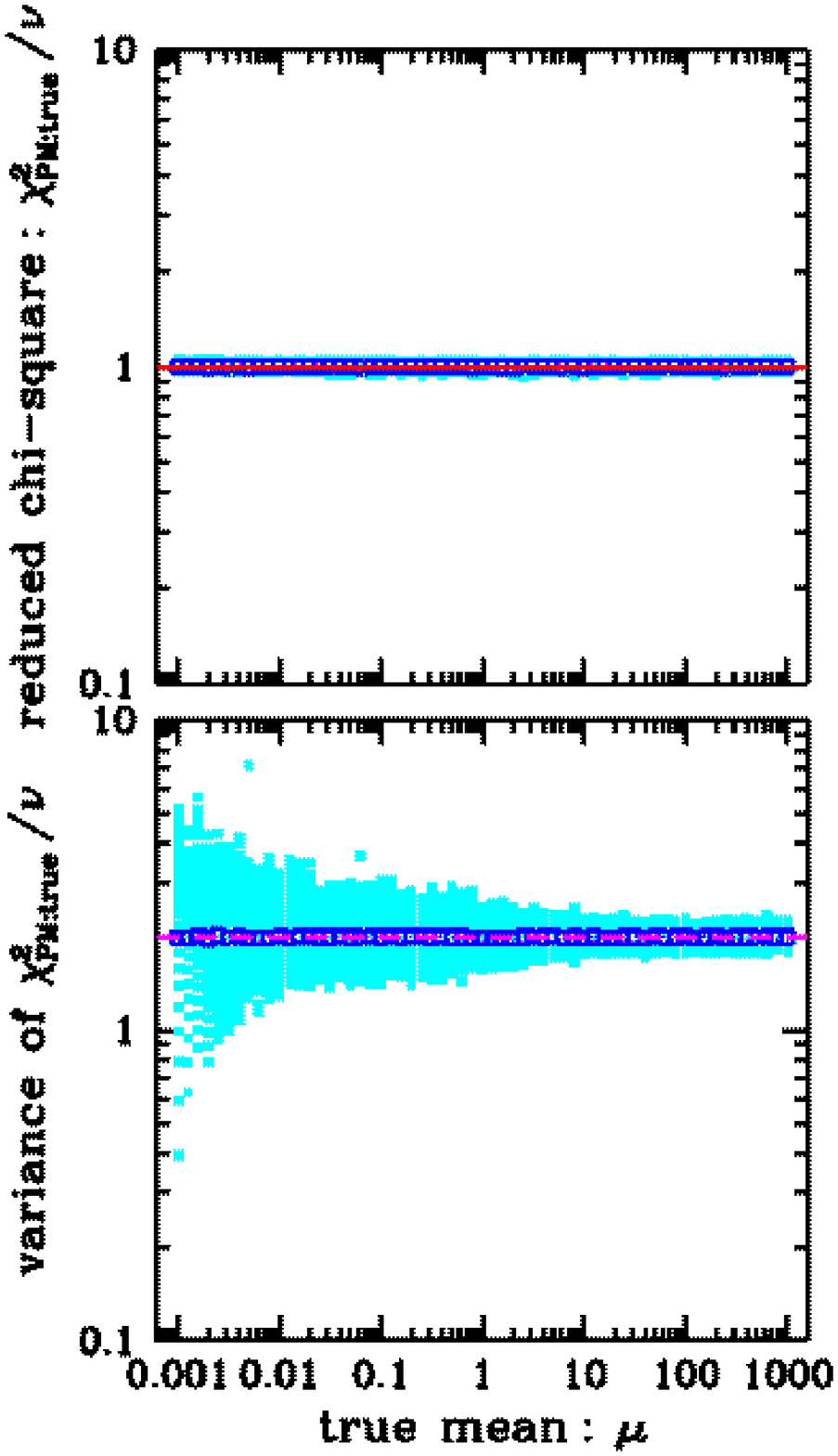}
  \vspace*{-40truemm}
  \caption[]{\baselineskip 1.15em \fig07cap}
  \end{figure}
}
\fi
\clearpage

\def\fig08cap{
\label{fig:x2pms_5mu}
\noteforeditor{Print this figure ONE (1) COLUMN wide.\newline}
The cumulative distribution functions for
1000 samples of $10^4$ Poisson deviates
({\em{top to bottom}}: $\mu \equiv 100, 10, 1, 0.1$, and 0.01)
analyzed using the modified Pearson's $\chi^2$ statistic
(same input data set as for Fig.\ \protect\ref{fig:x2l_5mu}).
In all cases, $\nu\!\equiv\!10^4$ and $m_i$ was set to the
{\em{sample mean}}.
Other details as in Fig.\ \protect\ref{fig:x2p_x2l_x2g_mu100}.
}

\ifundefined{showfigs}{
  \figcaption{\fig08cap}
}\else{
  \clearpage
  \newpage
  \begin{figure}
  \figurenum{8}
  \vspace*{-40truemm}
  \hspace*{+18truemm}
  \epsfxsize=6.5truein
  \epsfbox{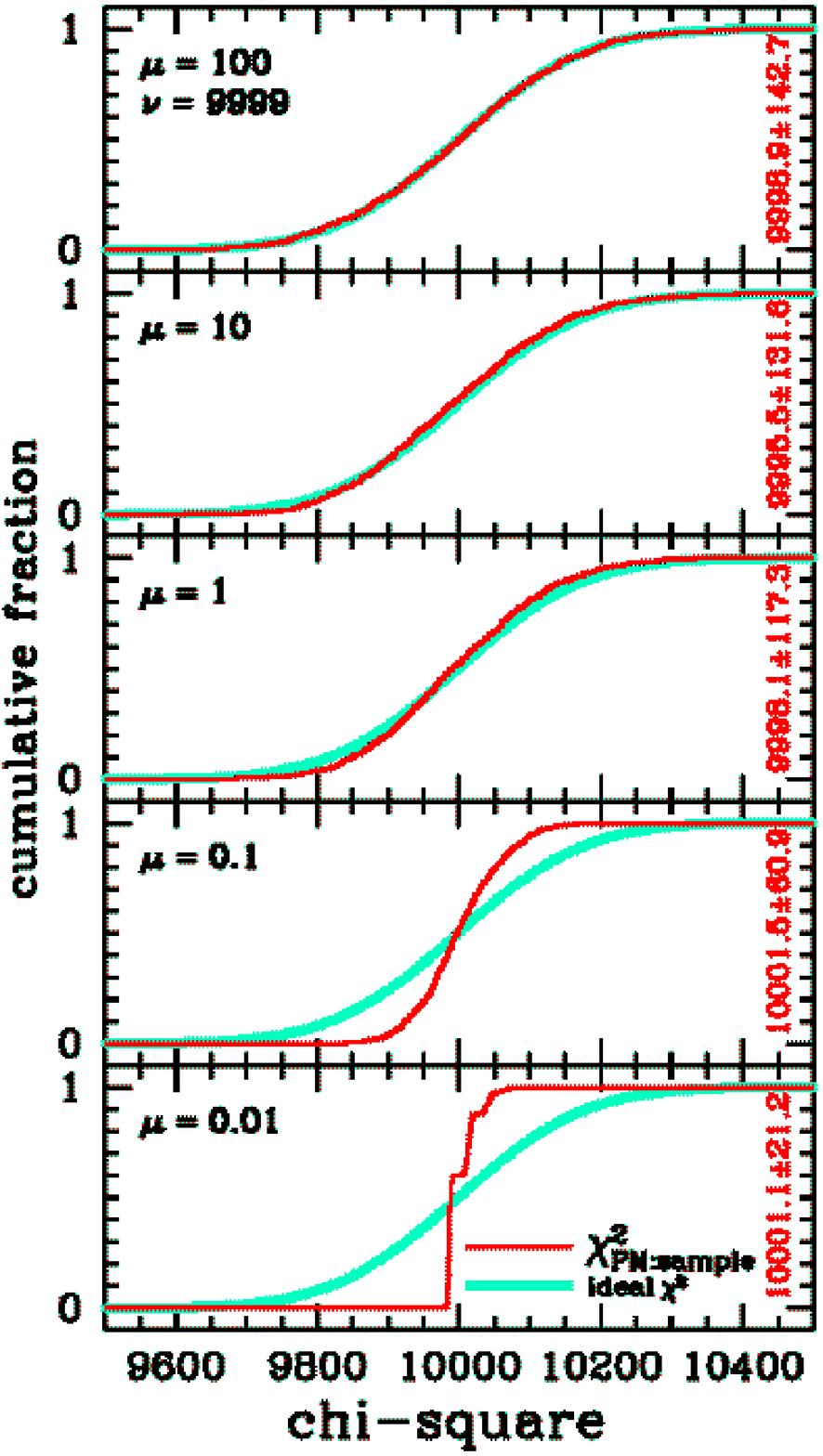}
  \vspace*{-40truemm}
  \caption[]{\baselineskip 1.15em \fig08cap}
  \end{figure}
}
\fi
\clearpage

\def\fig09cap{
\label{fig:x2pmsr_var}
\noteforeditor{Print this figure ONE (1) COLUMN wide.\newline}
Reduced chi-square as a function of the true Poisson mean for
the modified Pearson's $\chi^2$
statistic with the model of the $i$th deviate set to the
{\em{sample mean}}
(same input data set as for Fig.\ \protect\ref{fig:x2lr_var}).
Other details as in Fig.\ \protect\ref{fig:x2lr_var}.
}

\ifundefined{showfigs}{
  \figcaption{\fig09cap}
}\else{
  \clearpage
  \newpage
  \begin{figure}
  \figurenum{9}
  \vspace*{-40truemm}
  \hspace*{+18truemm}
  \epsfxsize=6.5truein
  \epsfbox{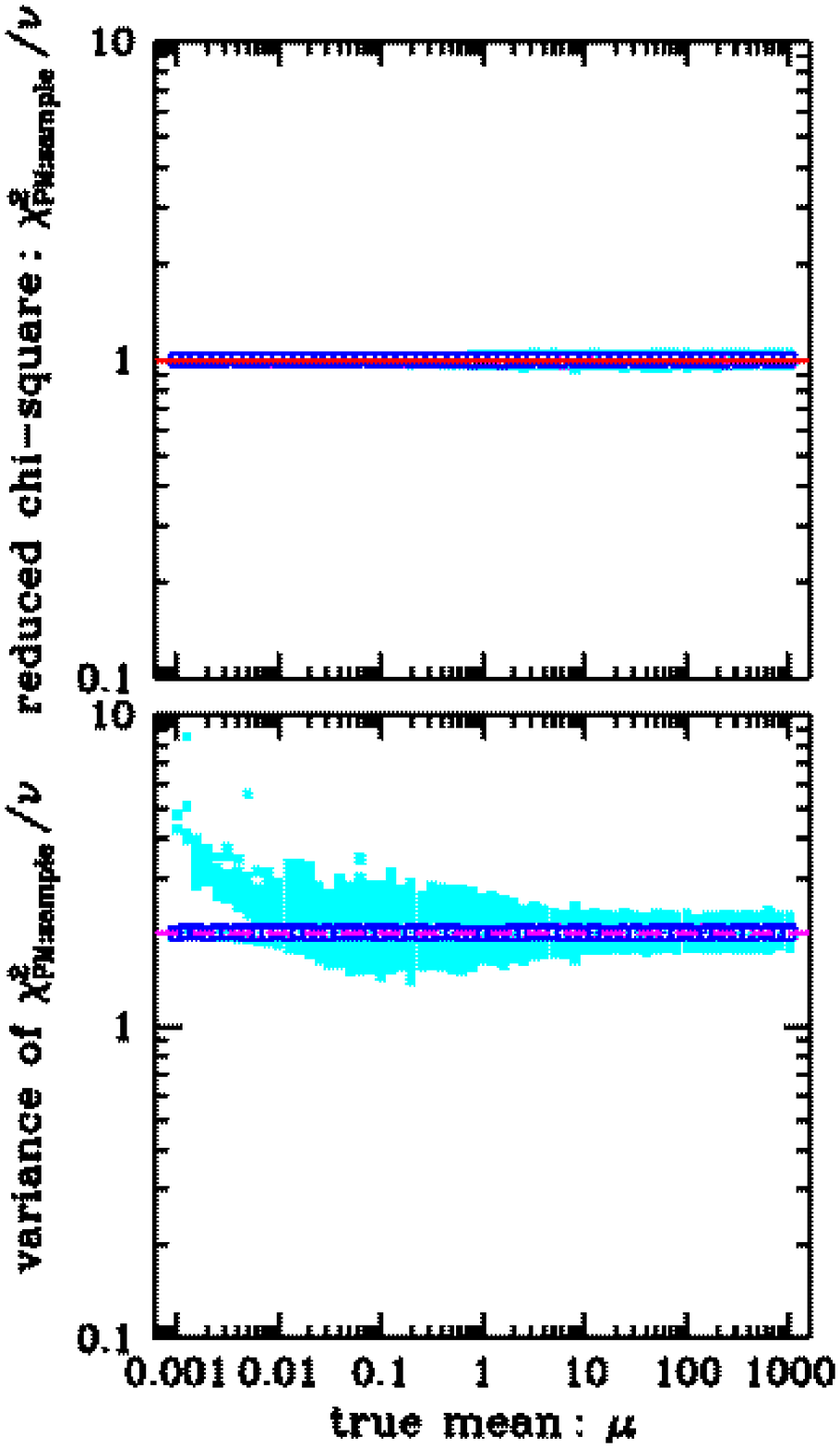}
  \vspace*{-40truemm}
  \caption[]{\baselineskip 1.15em \fig09cap}
  \end{figure}
}
\fi
\clearpage

\def\fig10cap{
\label{fig:x2gt_5mu}
\noteforeditor{Print this figure ONE (1) COLUMN wide.\newline}
The cumulative distribution functions for
1000 samples of $10^4$ Poisson deviates
({\em{top to bottom}}: $\mu \equiv 100, 10, 1, 0.1$, and 0.01)
analyzed using the $\chi^2_{\gamma}$ statistic
[definition:\ eq.\ (\ref{eq:x2g})]
(same input data set as for Fig.\ \protect\ref{fig:x2l_5mu}).
In all cases, $\nu\!\equiv\!10^4$ and $m_i$ was set to the true mean value
of the data set.
Other details as in Fig.\ \protect\ref{fig:x2p_x2l_x2g_mu100}.
}

\ifundefined{showfigs}{
  \figcaption{\fig10cap}
}\else{
  \clearpage
  \newpage
  \begin{figure}
  \figurenum{10}
  \vspace*{-40truemm}
  \hspace*{+18truemm}
  \epsfxsize=6.5truein
  \epsfbox{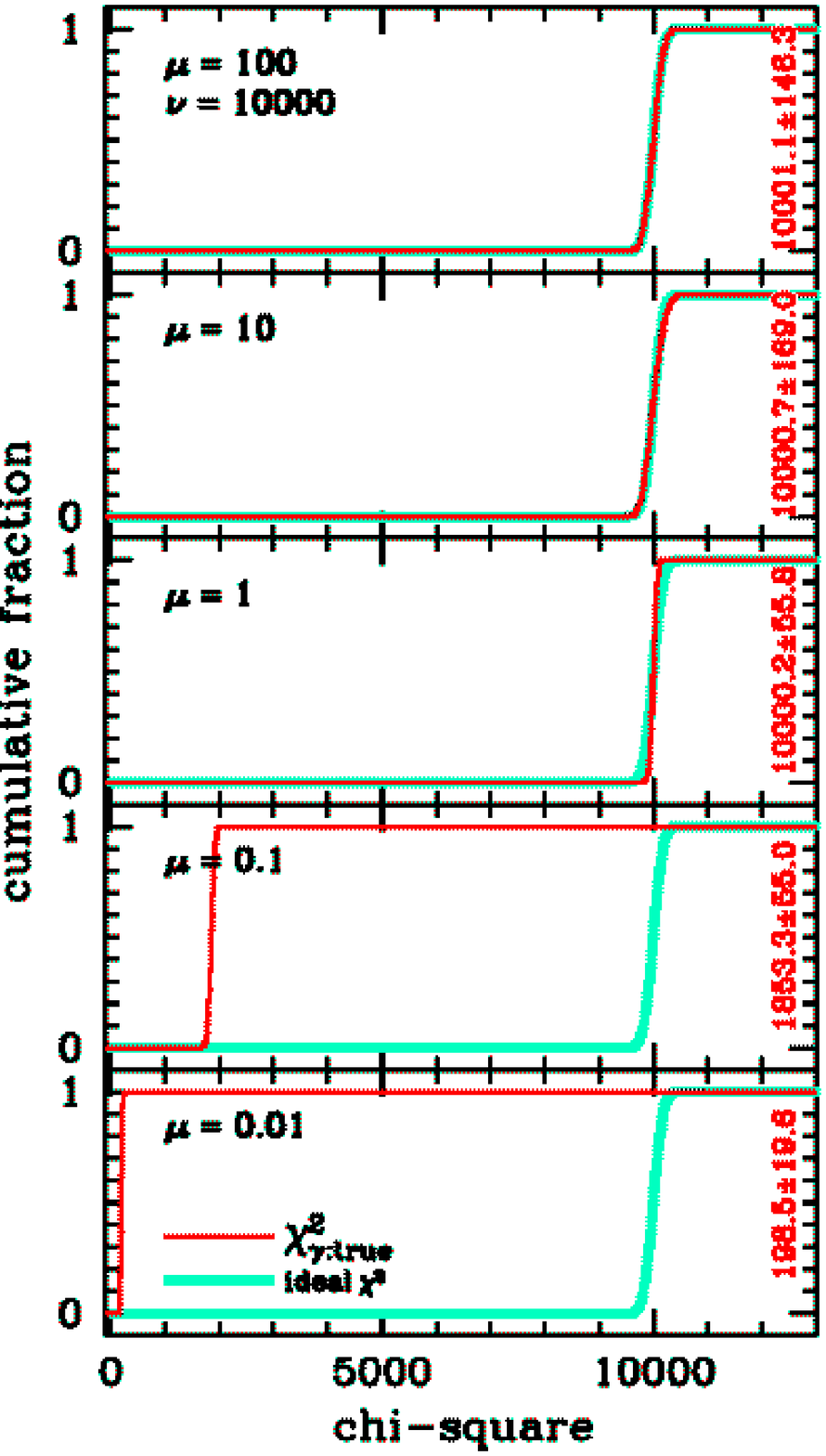}
  \vspace*{-40truemm}
  \caption[]{\baselineskip 1.15em \fig10cap}
  \end{figure}
}
\fi
\clearpage

\def\fig11cap{
\label{fig:x2gt_var}
\noteforeditor{Print this figure ONE (1) COLUMN wide.\newline}
Reduced chi-square as a function of the true Poisson mean for
the $\chi^2_{\gamma}$ statistic
statistic with the model of the $i$th deviate set to the
true mean value of the parent Poisson distribution
(same input data set as for Fig.\ \protect\ref{fig:x2lr_var}).
The {\em{solid line}} connecting the open squares in the
{\em{top panel}} is the formula
$1+e^{-\mu}\left(\mu-1\right)$
[eq.\ 29 of \paperi].
The {\em{solid line}} connecting the open squares in the
{\em{bottom panel}} is equation (\ref{eq:x2gr_var}).
Other details as in Fig.\ \protect\ref{fig:x2lr_var}.
}

\ifundefined{showfigs}{
  \figcaption{\fig11cap}
}\else{
  \clearpage
  \newpage
  \begin{figure}
  \figurenum{11}
  \vspace*{-40truemm}
  \hspace*{+18truemm}
  \epsfxsize=6.5truein
  \epsfbox{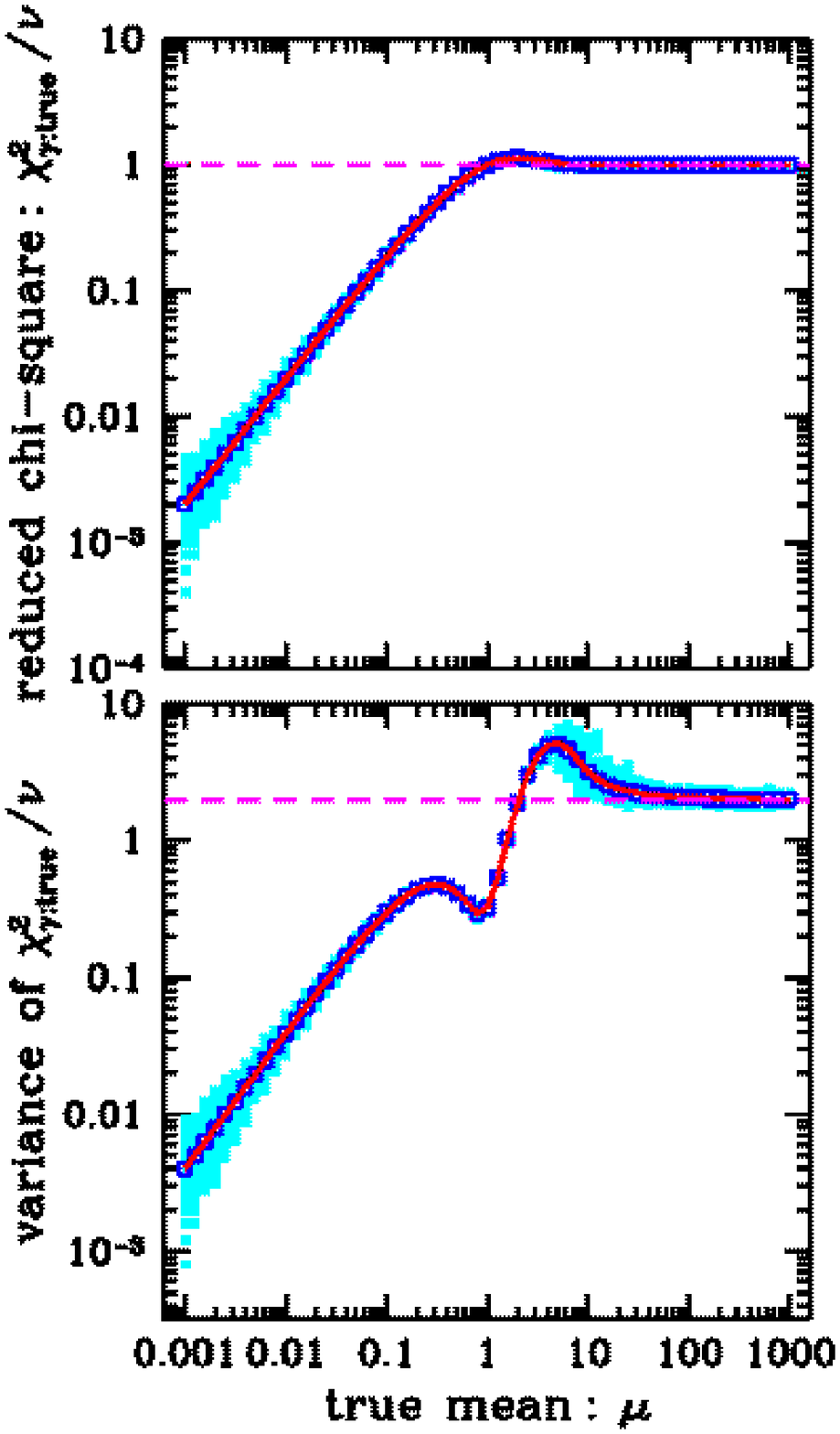}
  \vspace*{-40truemm}
  \caption[]{\baselineskip 1.15em \fig11cap}
  \end{figure}
}
\fi
\clearpage

\def\fig12cap{
\label{fig:x2gmt_5mu}
\noteforeditor{Print this figure ONE (1) COLUMN wide.\newline}
The cumulative distribution functions for
1000 samples of $10^4$ Poisson deviates
({\em{top to bottom}}: $\mu \equiv 100, 10, 1, 0.1$, and 0.01)
analyzed using the modified $\chi^2_{\gamma}$ statistic
[definition:\ eq.\ (\ref{eq:x2gm})]
(same input data set as for Fig.\ \protect\ref{fig:x2l_5mu}).
In all cases, $\nu\!\equiv\!10^4$ and $m_i$ was set to the true mean value
of the data set.
Other details as in Fig.\ \protect\ref{fig:x2p_x2l_x2g_mu100}.
}

\ifundefined{showfigs}{
  \figcaption{\fig12cap}
}\else{
  \clearpage
  \newpage
  \begin{figure}
  \figurenum{12}
  \vspace*{-40truemm}
  \hspace*{+18truemm}
  \epsfxsize=6.5truein
  \epsfbox{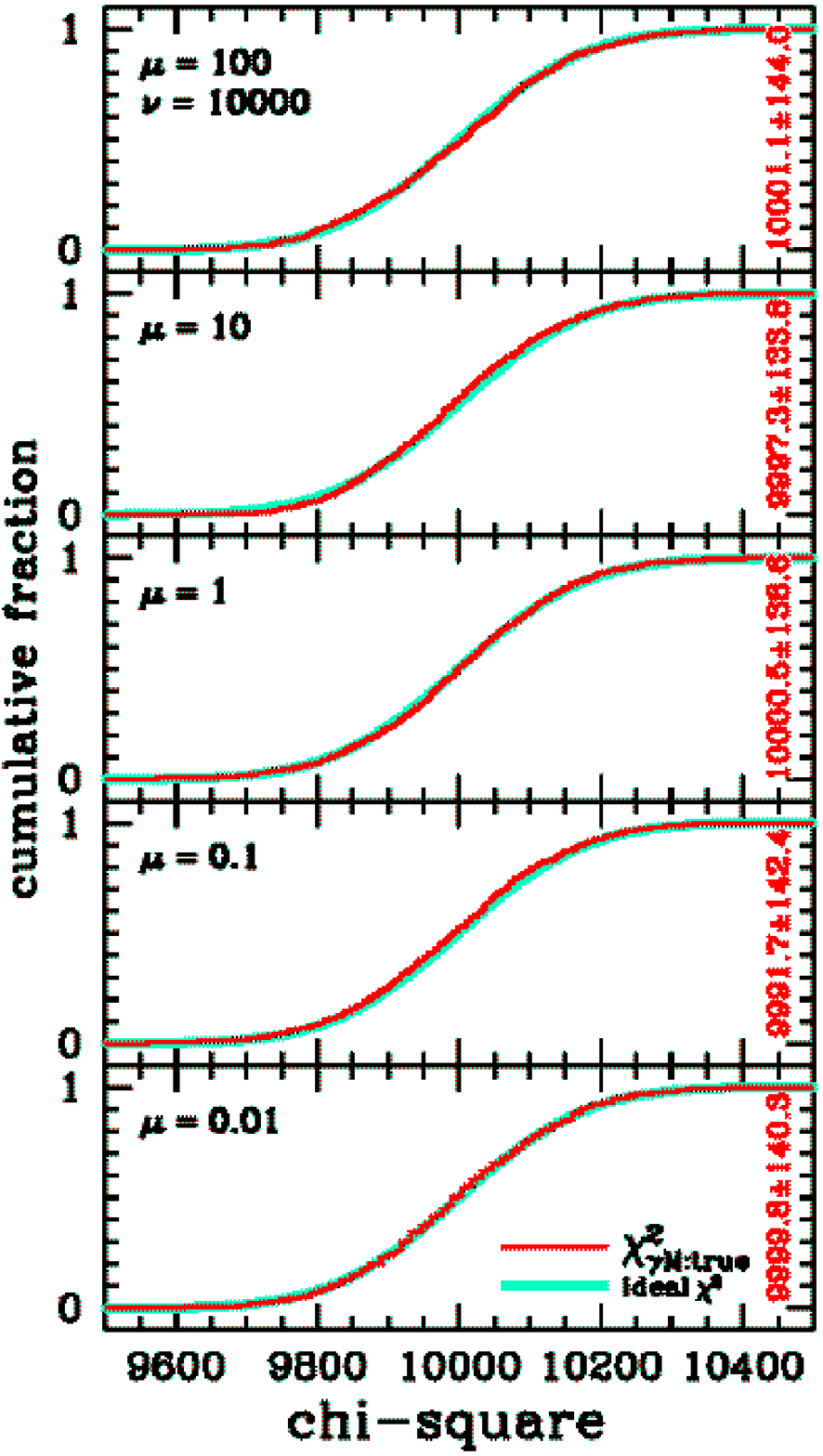}
  \vspace*{-40truemm}
  \caption[]{\baselineskip 1.15em \fig12cap}
  \end{figure}
}
\fi
\clearpage

\def\fig13cap{
\label{fig:x2gmt_var}
\noteforeditor{Print this figure ONE (1) COLUMN wide.\newline}
Reduced chi-square as a function of the true Poisson mean for
the modified $\chi^2_{\gamma}$ statistic
with the model of the $i$th deviate set to the
true mean value of the parent Poisson distribution
(same input data set as for Fig.\ \protect\ref{fig:x2lr_var}).
Other details as in Fig.\ \protect\ref{fig:x2lr_var}.
}

\ifundefined{showfigs}{
  \figcaption{\fig13cap}
}\else{
  \clearpage
  \newpage
  \begin{figure}
  \figurenum{13}
  \vspace*{-40truemm}
  \hspace*{+18truemm}
  \epsfxsize=6.5truein
  \epsfbox{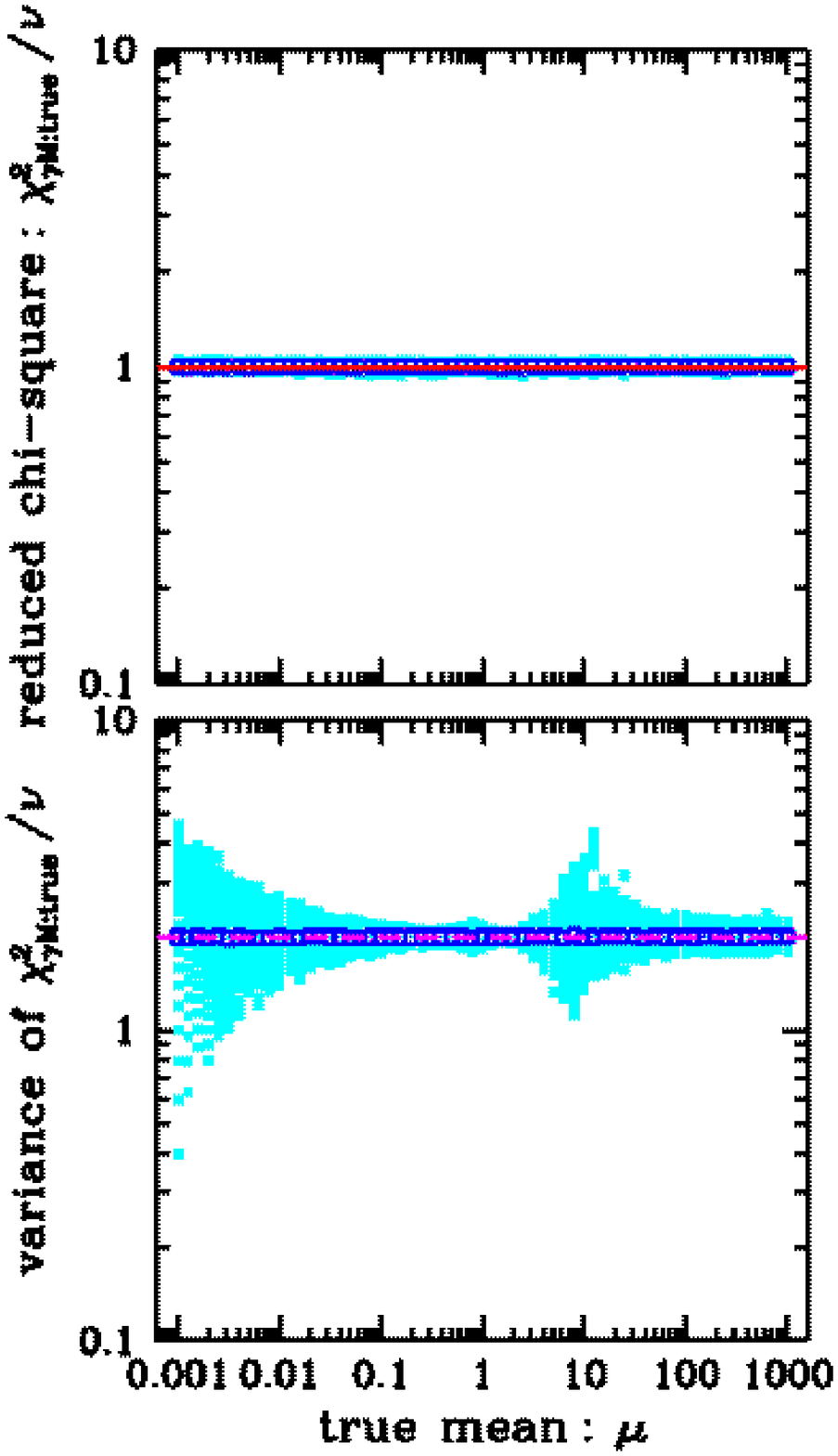}
  \vspace*{-40truemm}
  \caption[]{\baselineskip 1.15em \fig13cap}
  \end{figure}
}
\fi
\clearpage

\def\fig14cap{
\label{fig:x2gms_5mu}
\noteforeditor{Print this figure ONE (1) COLUMN wide.\newline}
The cumulative distribution functions for
1000 samples of $10^4$ Poisson deviates
({\em{top to bottom}}: $\mu \equiv 100, 10, 1, 0.1$, and 0.01)
analyzed using the modified $\chi^2_{\gamma}$ statistic
(same input data set as for Fig.\ \protect\ref{fig:x2l_5mu}).
In all cases, $\nu\!\equiv\!10^4$ and $m_i$ was set to the
{\em{sample mean}}.
Other details as in Fig.\ \protect\ref{fig:x2p_x2l_x2g_mu100}.
}

\ifundefined{showfigs}{
  \figcaption{\fig14cap}
}\else{
  \clearpage
  \newpage
  \begin{figure}
  \figurenum{14}
  \vspace*{-40truemm}
  \hspace*{+18truemm}
  \epsfxsize=6.5truein
  \epsfbox{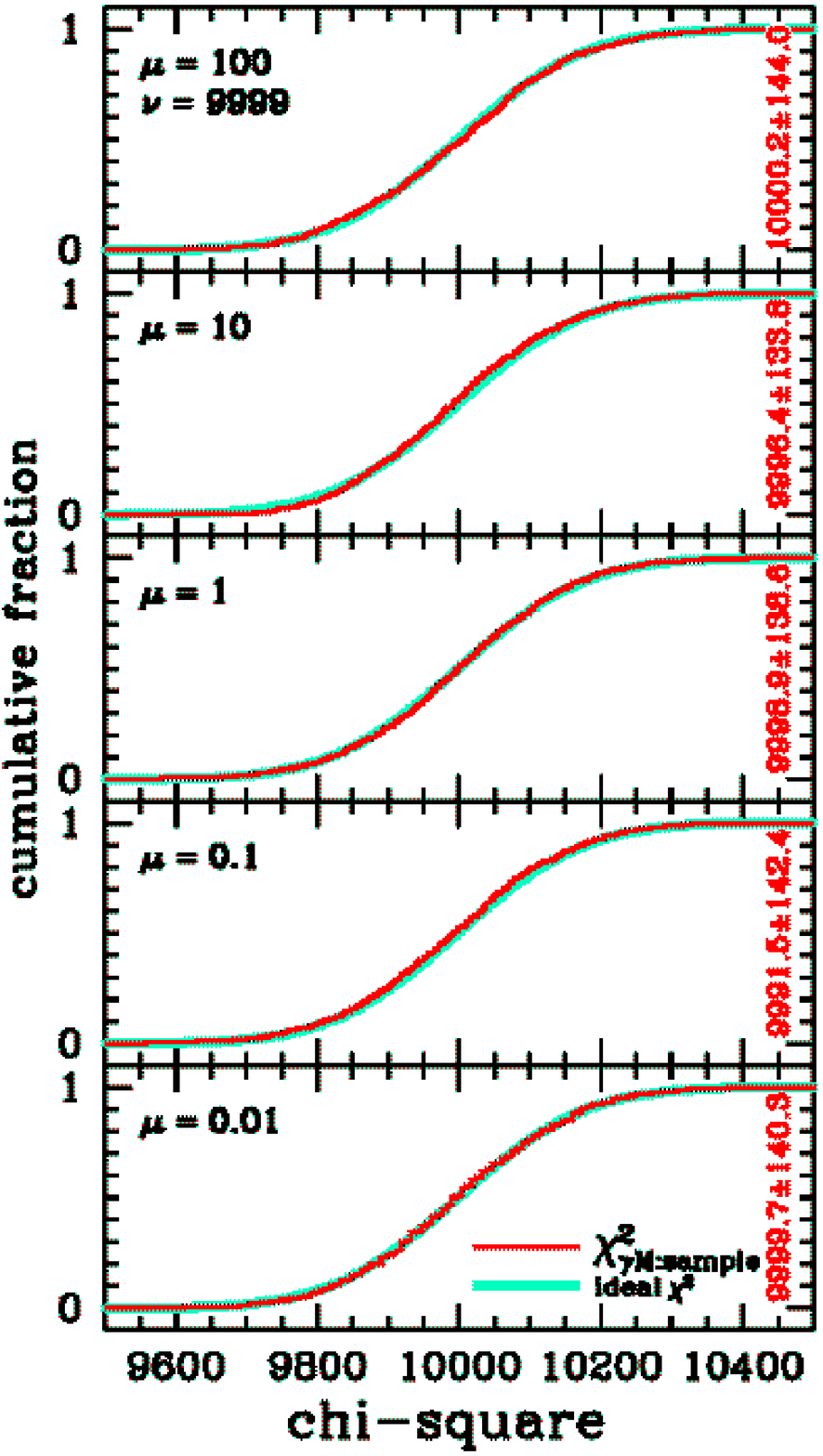}
  \vspace*{-40truemm}
  \caption[]{\baselineskip 1.15em \fig14cap}
  \end{figure}
}
\fi
\clearpage

\def\fig15cap{
\label{fig:x2gms_var}
\noteforeditor{Print this figure ONE (1) COLUMN wide.\newline}
Reduced chi-square as a function of the true Poisson mean for
the modified $\chi^2_{\gamma}$ statistic
with the model of the $i$th deviate set to the
{\em{sample mean}}
(same input data set as for Fig.\ \protect\ref{fig:x2lr_var}).
Other details as in Fig.\ \protect\ref{fig:x2lr_var}.
}

\ifundefined{showfigs}{
  \figcaption{\fig15cap}
}\else{
  \clearpage
  \newpage
  \begin{figure}
  \figurenum{15}
  \vspace*{-40truemm}
  \hspace*{+18truemm}
  \epsfxsize=6.5truein
  \epsfbox{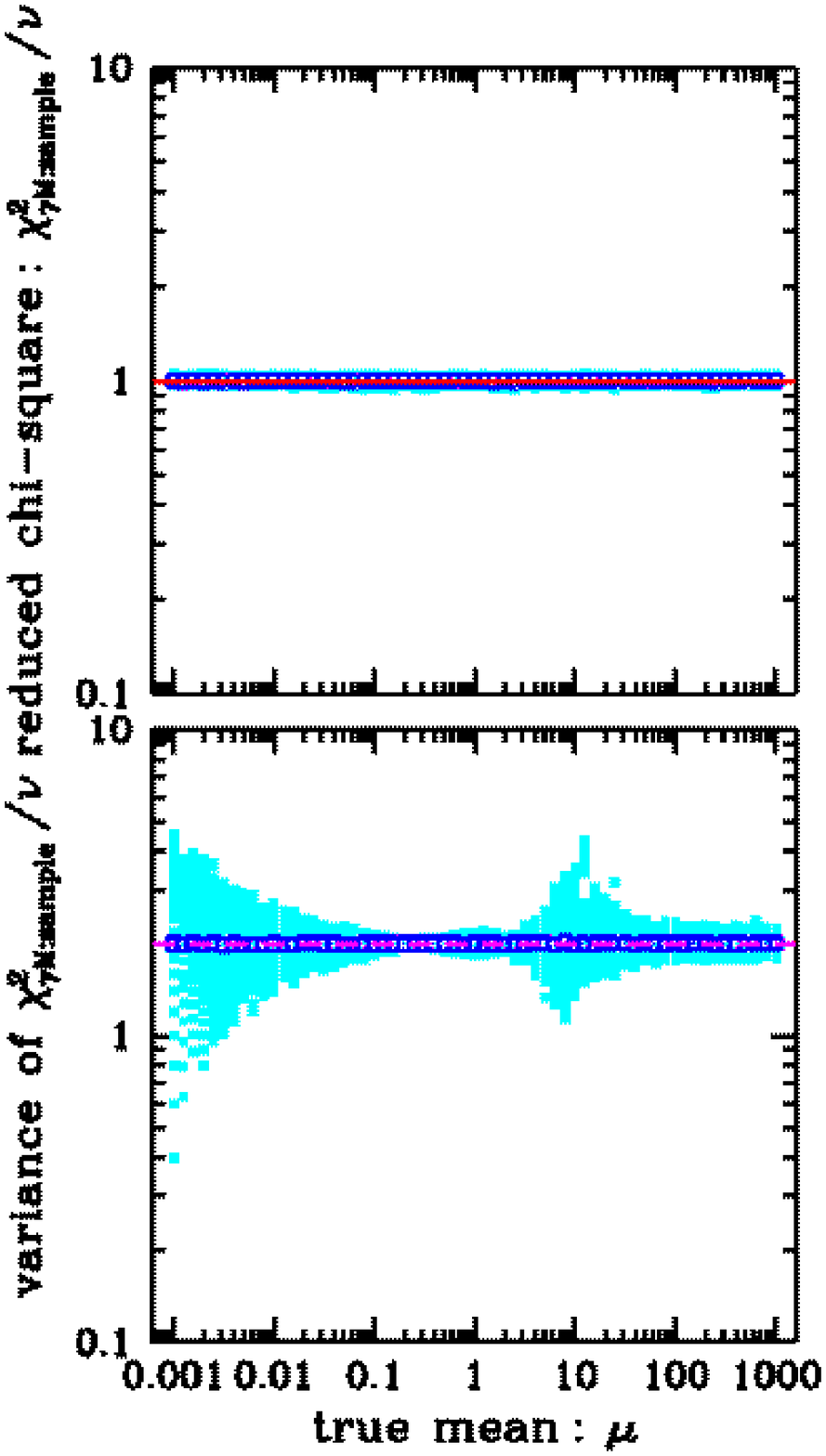}
  \vspace*{-40truemm}
  \caption[]{\baselineskip 1.15em \fig15cap}
  \end{figure}
}
\fi
\clearpage

\def\fig16cap{
\label{fig:observation}
\noteforeditor{Print this figure TWO (2) COLUMNS wide.\newline}
The simulated X-ray observation.
A 40 photon X-ray point source is located
at the $(x,y)$ position of $(33,26)$ on a
background of 0.06 photons per pixel.
The Point Spread Function is
$\phi(x,y) \equiv(\pi/3)[1-\mbox{min}(r,1)]$
where $r^2 = (x/10)^2 + (y/10)^2$.
This is the same PSF used
by Cash (\protect\cite{ca1979}) but with
a resolution of 100 pixels per unit area.
}

\ifundefined{showfigs}{
  \figcaption{\fig16cap}
}\else{
  \clearpage
  \newpage
  \begin{figure}
  \figurenum{16}
  \vspace*{-45truemm}
  \hspace*{+0truemm}
  \epsfxsize=6.5truein
  \epsfbox{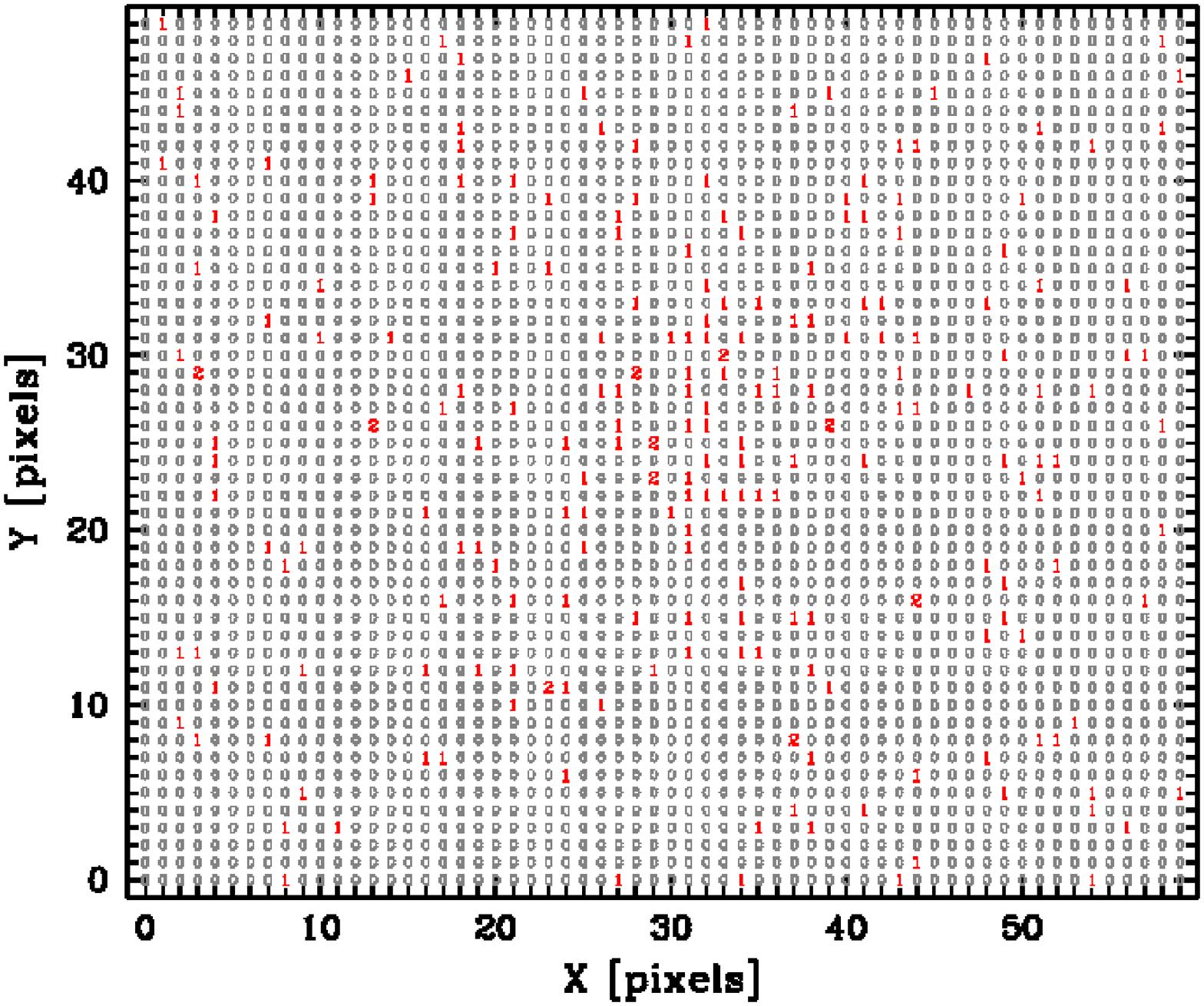}
  \vspace*{-35truemm}
  \caption[]{\baselineskip 1.15em \fig16cap}
  \end{figure}
}
\fi
\clearpage

\def\fig17cap{
\label{fig:first_sky}
\noteforeditor{Print this figure TWO (2) COLUMNS wide.\newline}
{\em Gray boxes}
indicate pixels with a total number of photons within a radius of 10 pixels
that are consistent
(within the 99.9\% upper and lower single-sided confidence limits of the
observed photon total)
with the estimated background flux level of
0.0747 photons per pixel.
The {\em X} marks the center of the X-ray point source and the
{\em dotted circle} has a radius of 10 pixels which is the
maximum size of the PSF.
Other details as in Fig.\ \protect\ref{fig:observation}.
}

\ifundefined{showfigs}{
  \figcaption{\fig17cap}
}\else{
  \clearpage
  \newpage
  \begin{figure}
  \figurenum{17}
  \vspace*{-45truemm}
  \hspace*{+0truemm}
  \epsfxsize=6.5truein
  \epsfbox{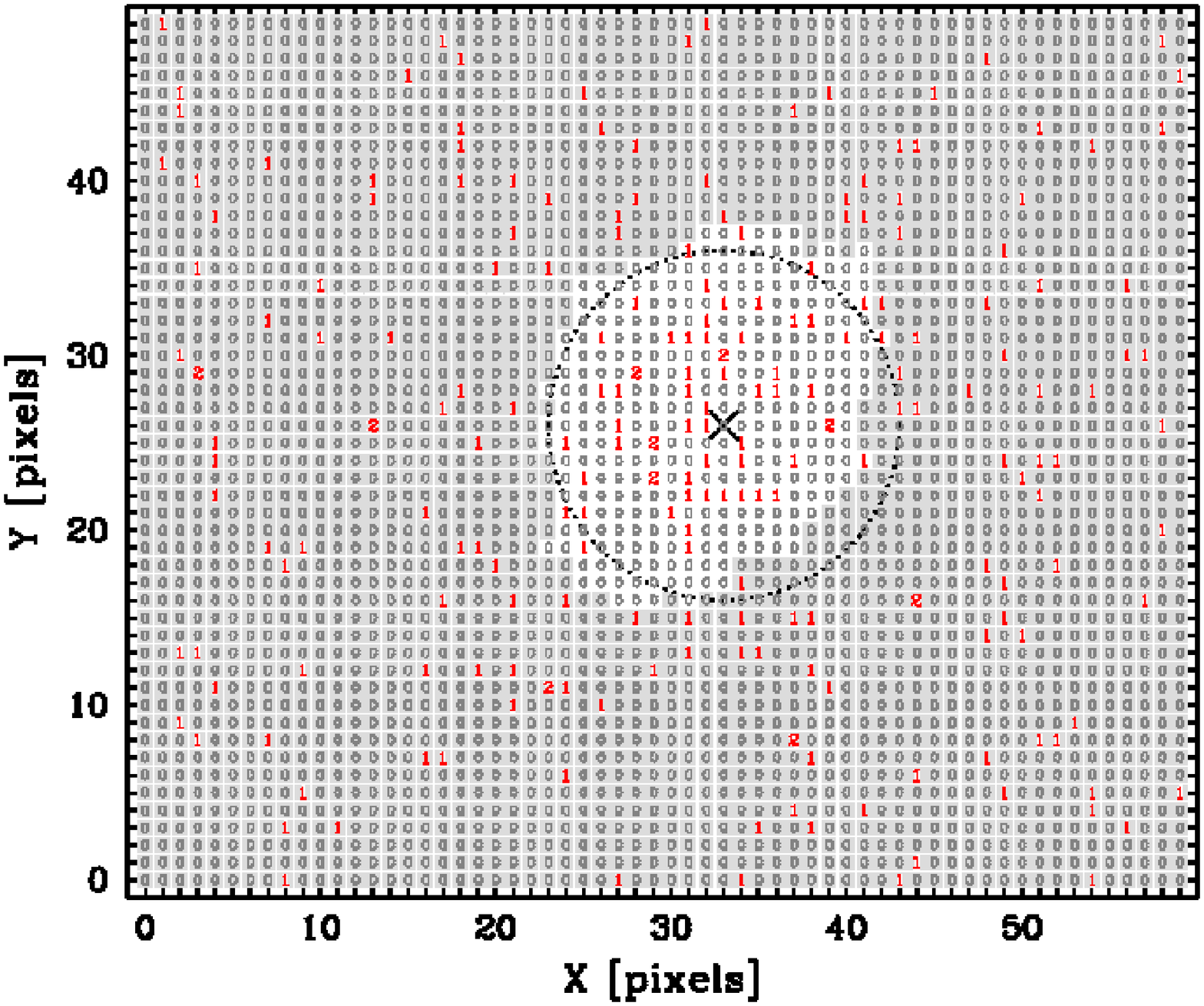}
  \vspace*{-35truemm}
  \caption[]{\baselineskip 1.15em \fig17cap}
  \end{figure}
}
\fi
\clearpage

\def\fig18cap{
\label{fig:final_sky}
\noteforeditor{Print this figure TWO (2) COLUMNS wide.\newline}
{\em Gray boxes} indicate the
pixels with a total number of photons within a radius of 10 pixels
that are consistent
(within the 99.9\% upper and lower single-sided confidence limits of the
observed photon total)
with the true background flux level of
\underline{0.06} photons per pixel.
{\em Dark-gray circles} indicate the
pixels with a total number of photons within a radius of 10 pixels
that are consistent
(within the \underline{95\%}
upper and lower single-sided confidence limits of the
observed photon total)
with the model of a 40 photon point source {\em at that pixel location}
on a background of 0.06 photons per pixel.
Other details as in Fig.\ \protect\ref{fig:first_sky}.
}

\ifundefined{showfigs}{
  \figcaption{\fig18cap}
}\else{
  \clearpage
  \newpage
  \begin{figure}
  \figurenum{18}
  \vspace*{-45truemm}
  \hspace*{+0truemm}
  \epsfxsize=6.5truein
  \epsfbox{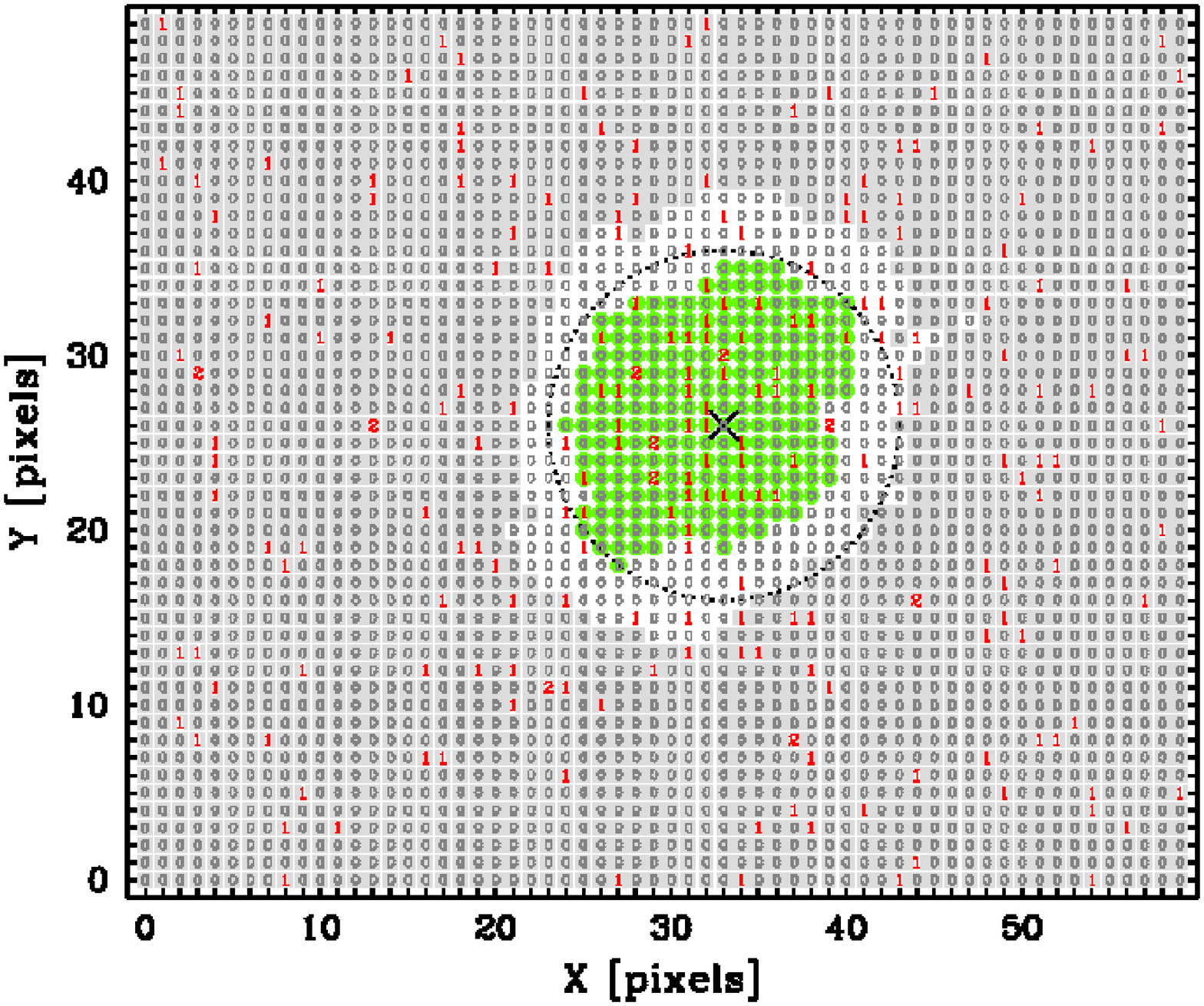}
  \vspace*{-35truemm}
  \caption[]{\baselineskip 1.15em \fig18cap}
  \end{figure}
}
\fi
\clearpage

\def\fig19cap{
\label{fig:green_radius8}
\noteforeditor{Print this figure TWO (2) COLUMNS wide.\newline}
{\em Dark-gray circles} indicate the
pixels with a total number of photons within a radius of \underline{8 pixels}
that are consistent
(within the 95\% upper and lower single-sided confidence limits of the
observed photon total)
with the model of a
40 photon point source at that pixel location
on a background of 0.06 photons per pixel.
All pixels within the {\em solid black contour}
are within the \underline{95\% confidence interval} as
determined by the $\chi^2_{\gamma{\rm M}}$
analysis of the cumulative radial distribution of the data
within 10 pixels is compared with the cumulative radial distribution
of a model of a 40 photon point source at that pixel location
on a background of 0.06 photons per pixel.
Note how well the 95\% confidence interval of the
$\chi^2_{\gamma{\rm M}}$ analysis of the cumulative radial distribution
matches the region described by the circled pixels.
Other details as in Fig.\ \protect\ref{fig:final_sky}.
}

\ifundefined{showfigs}{
  \figcaption{\fig19cap}
}\else{
  \clearpage
  \newpage
  \begin{figure}
  \figurenum{19}
  \vspace*{-45truemm}
  \hspace*{+0truemm}
  \epsfxsize=6.5truein
  \epsfbox{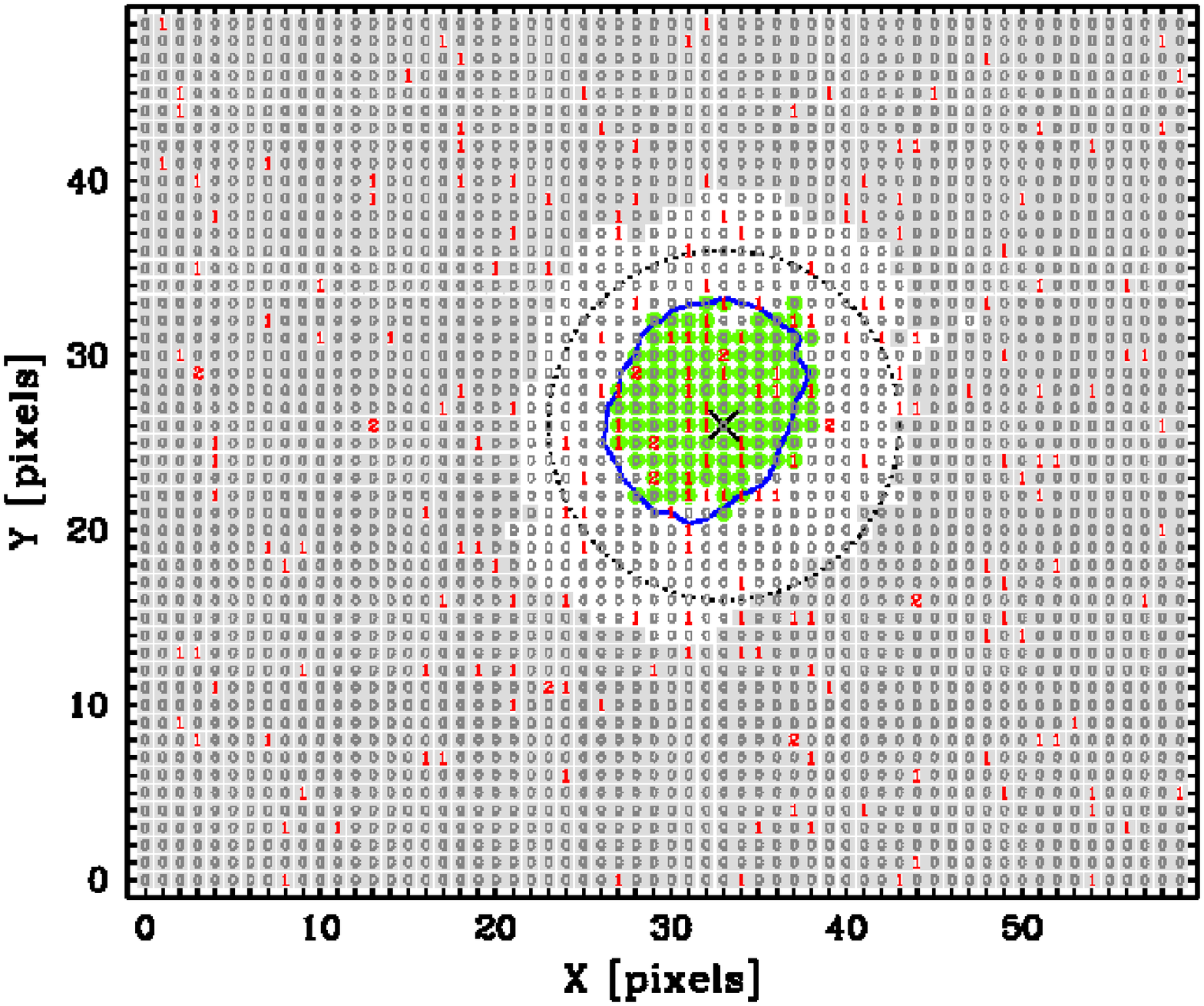}
  \vspace*{-35truemm}
  \caption[]{\baselineskip 1.15em \fig19cap}
  \end{figure}
}
\fi
\clearpage

\def\fig20cap{
\label{fig:crd_probability}
\noteforeditor{Print this figure TWO (2) COLUMNS wide.\newline}
The photon distribution of the 317 pixels within a radius of 10 pixels
of the location $(33,26)$ of Fig.\ \protect\ref{fig:observation}
was transformed to a cumulative radial distribution
(10 1-pixel-wide bins $\Rightarrow$ 10 degrees-of-freedom)
and then compared,
using the modified chi-square-gamma statistic,
with 80 models of the observation:
a 1 to 80 photon (in steps of 1 photon) X-ray point source at $(33,26)$
on a background of 0.06 photons per pixel (i.e., the true background).
The 95th percentage point for the chi-square distribution with 10
degrees of freedom is 18.31 [i.e. $P(18.31|10)=0.95\,$].
Assuming that $\chi^2_{\gamma{\rm M}}$ is distributed like $\chi^2$,
we see that the
upper and lower single-sided 95\% confidence limits for
the intensity of an X-ray point source at $(33,26)$ in
Fig.\ \protect\ref{fig:observation} is 54.5 and 28.0, respectively.
The true intensity of the X-ray source is 40 photons.
}

\ifundefined{showfigs}{
  \figcaption{\fig20cap}
}\else{
  \clearpage
  \newpage
  \begin{figure}
  \figurenum{20}
  \vspace*{-45truemm}
  \hspace*{+0truemm}
  \epsfxsize=6.5truein
  \epsfbox{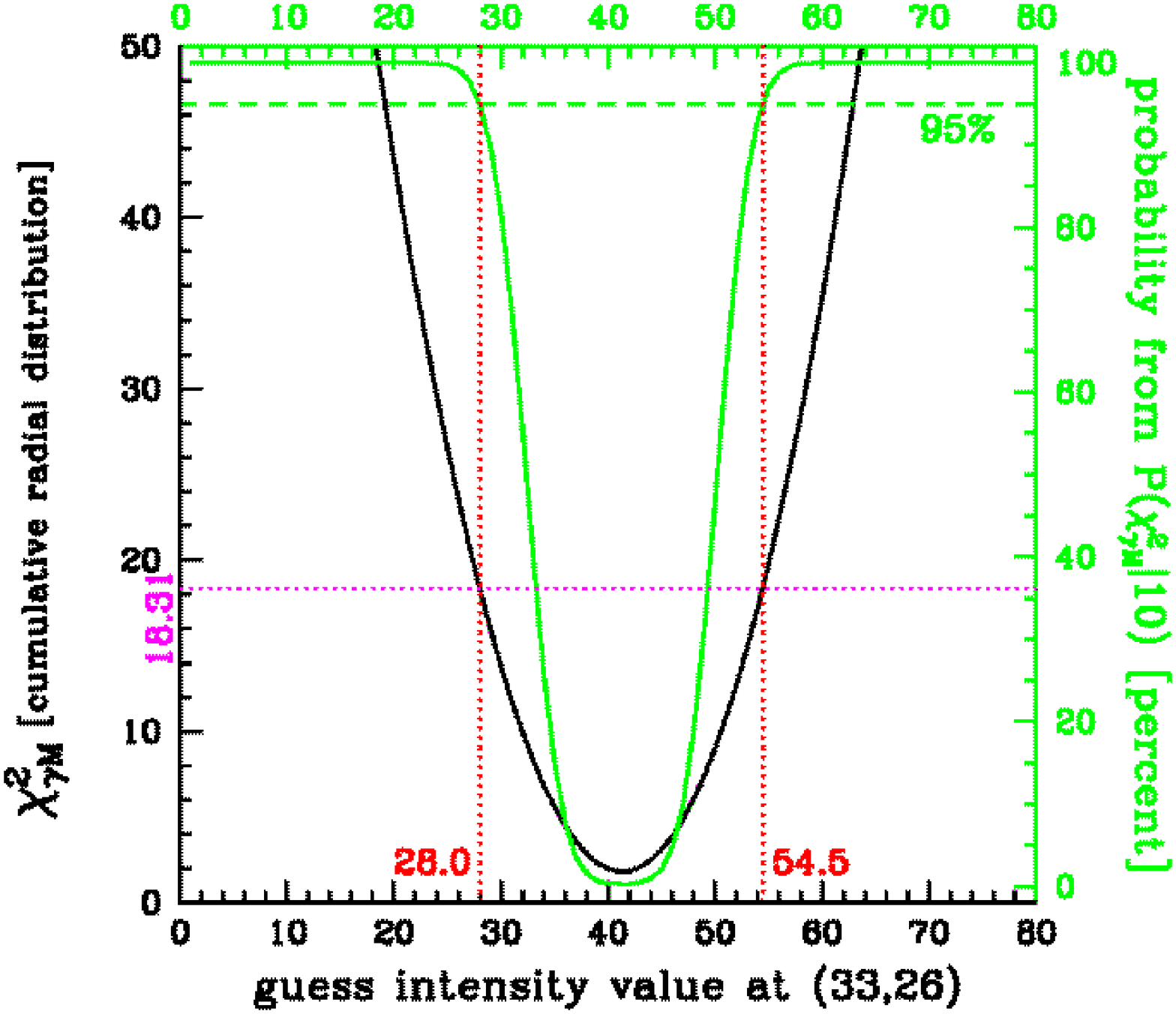}
  \vspace*{-35truemm}
  \caption[]{\baselineskip 1.15em \fig20cap}
  \end{figure}
}
\fi
\clearpage

\def\fig21cap{
\label{fig:10000_cumulative_fraction}
\noteforeditor{Print this figure ONE (1) COLUMN wide.\newline}
A data set of $10^4$ realizations of the
the same model used to make Fig.\ \protect\ref{fig:observation}
was created.
Each sample in this data set was then analyzed using
the modified chi-square-gamma statistic
at the location $(33,26)$ -- the true location of the simulated
X-ray point source of 40 photons on a background of 0.06 photons per pixel.
All 317 pixels within a radius of 10 pixels
(the size of the PSF) were compared to the
true model value at that location
and
the number of independent degrees-of-freedom was
therefore equal to the number of pixels analyzed
(i.e.\ $\nu\!\equiv\!317$).
Compare the cumulative distribution
with the cumulative distribution function of a Gaussian distribution
with a mean of $317$ and a variance of $\sqrt{2\!\times\!317}~(\approx25.2)$
[{\em{thick curve}}].
The numbers with errors shown on the right side give the
mean and rms value for the
$\chi^2_{\gamma{\rm M}}$ ({\em top})
and
ideal $\chi^2$ ({\em bottom})
statistics.
}

\ifundefined{showfigs}{
  \figcaption{\fig21cap}
}\else{
  \clearpage
  \newpage
  \begin{figure}
  \figurenum{21}
  \vspace*{-38truemm}
  \hspace*{+0truemm}
  \epsfxsize=6.5truein
  \epsfbox{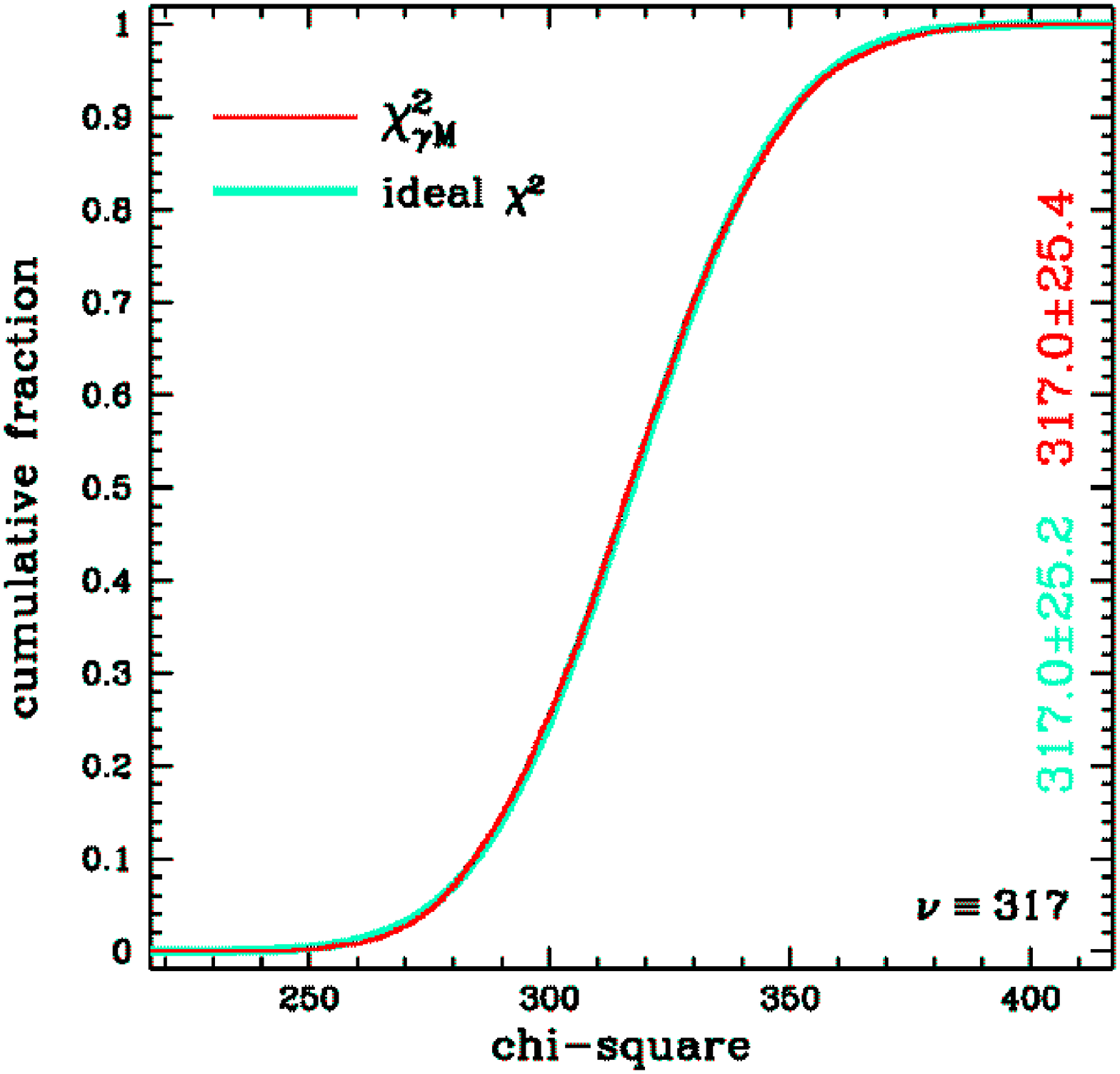}
  \vspace*{-40truemm}
  \caption[]{\baselineskip 1.15em \fig21cap}
  \end{figure}
}
\fi
\clearpage

\def\fig22cap{
\label{fig:10000_sorted_probability}
\noteforeditor{Print this figure ONE (1) COLUMN wide.\newline}
The $10^4$ simulations
of Fig.\ \protect\ref{fig:10000_cumulative_fraction}
were sorted by the value of the
modified chi-square-gamma statistic.
The {\em dark} plot shows the probablity $P(\chi^2_{\gamma{\rm M}}|317)$
as a function of the sorted $\chi^2_{\gamma{\rm M}}$ values.
The {\em gray} plot on the bottom shows the residuals from the ideal
one-to-one correspondance.
The {\em predicted}
probabilities for the 9000th, 9500th, and 9900th sorted simulations
were 90.2069\%, 94.8666\%, and 98.9497\%, which agrees very well with
the {\em expected} probabilities of 90\%, 95\%, and 99\%, respectively.
For the entire simulation, the
mean and rms value of the resiuals is $0.0013\!\pm\!0.0038$ percentage
points --- note that the residuals never exceeds 1 percent.
Figures
\protect\ref{fig:10000_cumulative_fraction}
and
\protect\ref{fig:10000_sorted_probability}
indicate that the assumption that
the $\chi^2_{\gamma{\rm M}}$ statistic
is distributed like the $\chi^2$ distribution was valid.
}

\ifundefined{showfigs}{
  \figcaption{\fig22cap}
}\else{
  \clearpage
  \newpage
  \begin{figure}
  \figurenum{22}
  \vspace*{-38truemm}
  \hspace*{+0truemm}
  \epsfxsize=6.5truein
  \epsfbox{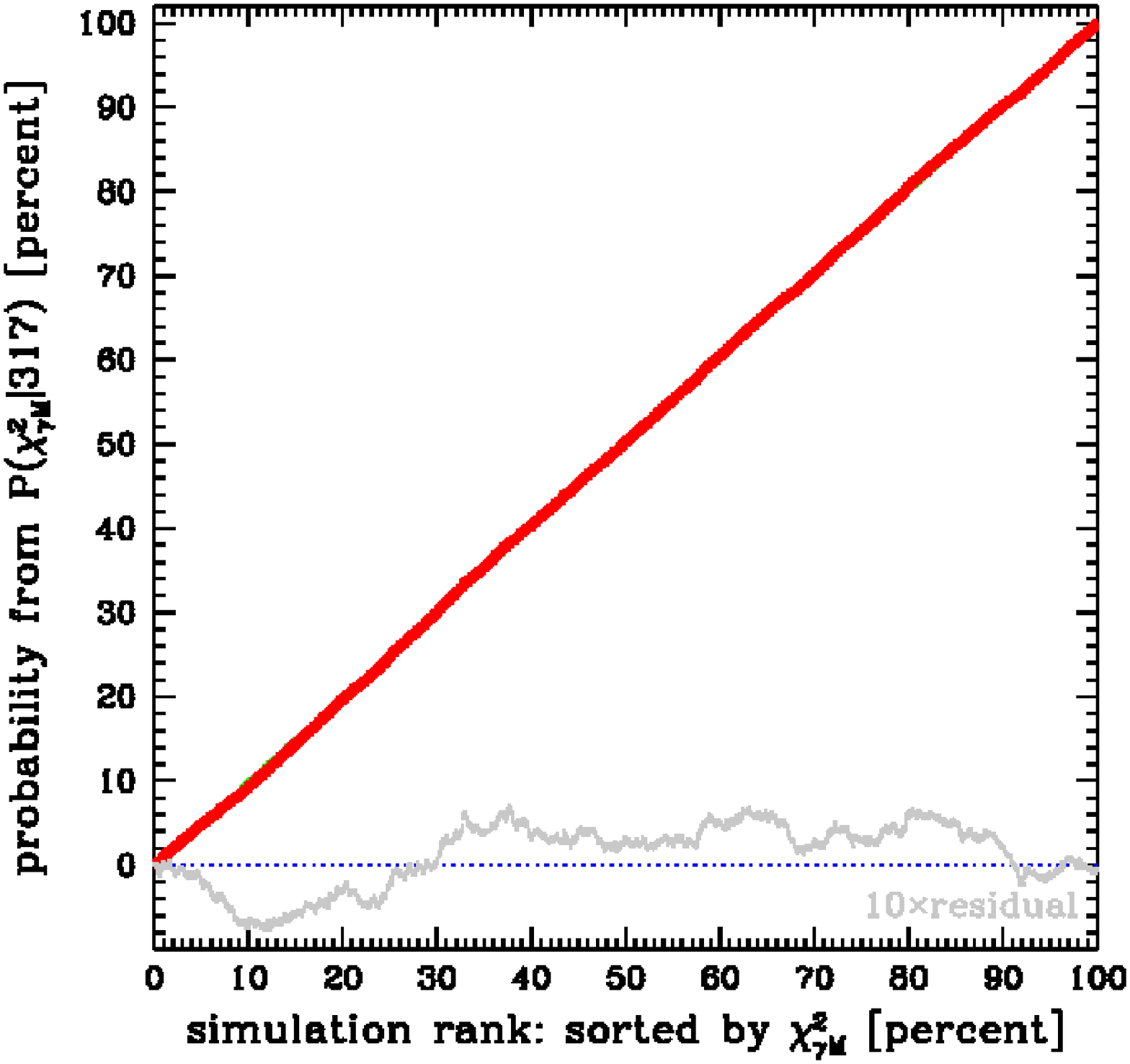}
  \vspace*{-40truemm}
  \caption[]{\baselineskip 1.15em \fig22cap}
  \end{figure}
}
\fi
\clearpage

\end{document}